\documentclass[12pt]{article}

\usepackage[left=1in,top=1in,right=1in,bottom=1in]{geometry}
\usepackage{amssymb}
\usepackage{amsmath}
\usepackage{bm}

\usepackage{graphicx}        

\usepackage{stmaryrd}
\SetSymbolFont{stmry}{bold}{U}{stmry}{m}{n}

\newtheorem{theorem}{Theorem}
\newtheorem{remark}{Remark}

\begin{document}
\title{Thermal convection in the van der Waals fluid}
\author{Ju Liu \\
\textit{\small Department of Pediatrics (Cardiology),}\\
\textit{\small Institute for Computational \& Mathematical Engineering, Stanford University}\\
\textit{\small Clark Center E1.3, 318 Campus Drive, Stanford, CA 94305, USA}\\
\small \textit{E-mail address:} liuju@stanford.edu, liujuy@gmail.com
}

\date{}
%
%
\maketitle

\section*{Abstract}
In this work, the van der Waals fluid model, a diffuse-interface model for liquid-vapor two-phase flows, is numerically investigated. The thermodynamic properties of the van der Waals fluid are first studied. Dimensional analysis is performed to identify the control parameters for the system. An entropy-stable numerical scheme and isogeometric analysis are utilized to discretize the governing equations for numerical simulations. The steady state solution at low Rayleigh number is presented, demonstrating the capability of the model in describing liquid-vapor phase transitions. Next, two-dimensional nucleate and film boiling are simulated, showing the applicability of the model in different regimes of boiling. In the last, the heat transport property of the van der Waals model is numerically investigated. The scaling law for the Nusselt number with respect to the Rayleigh number in the van der Waals model is obtained by performing a suite of high-resolution simulations.\\

\noindent \textbf{Keywords:} Boiling heat transfer, Van der Waals fluid, Rayleigh-B\'enard convection, Non-Oberbeck-Boussinesq effect

\section{Introduction}
The Rayleigh-B\'enard convection describes the buoyancy driven flow confined by a hot bottom plate and a cold top plate. It is a canonical model for studying hydrodynamic stability \cite{Chandrasekhar1981} and thermally driven turbulence \cite{Ahlers2009}. The bulk transport properties for the Rayleigh-B\'enard convective flow is mysterious and still attracts a tremendous amount of scientific interests world-widely. Traditionally, the Oberbeck-Boussinesq (OB) approximation for buoyancy has been utilized as a standard mathematical model to investigate the Rayleigh-B\'enard instability \cite{Drazin1981}. In the OB approximation, it is assumed that the fluid density varies linearly with the temperature field, the flow is incompressible, and the heat produced by internal friction is negligible. Based on the OB approximation, various scaling theories have been developed to account for the relationship between the bulk fluid transport properties (e.g. the Nusselt number $\operatorname{Nu}$, the Reynolds number $\operatorname{Re}$, etc.) and the strength of the buoyant force, which is measured by the Rayleigh number $\operatorname{Ra}$. Theoretical, experimental, and numerical studies suggest that $\operatorname{Nu} \propto \operatorname{Ra}^{\gamma_{\textup{Nu}}}$ and $\operatorname{Re} \propto \operatorname{Ra}^{\gamma_{\textup{Re}}}$ \cite{Ahlers2009}. Currently, the precise values of $\gamma_{\textup{Nu}}$ and $\gamma_{\textup{Re}}$ at different fluid regimes still remain under debate, and the existence of the theoretically predicted `ultimate state' at high Rayleigh number still awaits experimental and numerical confirmation \cite{Sommeria1999}.

Recent research activity has moved beyond the classical Oberbeck-Boussinesq model \cite{Xia2013}, and investigations of the non-Oberbeck-Boussinesq (NOB) effects have been carried out \cite{Accary2008,Oresta2009}. In more practical situations, fluid properties can be temperature dependent \cite{Tackley1996}; fluids may experience compressibility effects or even phase transition. In the literature, numerical investigations of the convection in ideal fluids and super-critical fluids have been performed \cite{Accary2008,Furukawa2002,Tilgner2011}. A recent research work has investigated the convection in multiphase fluids \cite{Xia2013}. In \cite{Lakkaraju2013,Oresta2009}, researchers modified the classical OB model by introducing point sources in the balance equations to model the behavior of bubbles in boiling flows. This approach necessitates empirical knowledge of several physical coefficients related to the bubbles. The lattice Boltzmann model has been utilized in combination with non-ideal gas models to study heat convection in two-phase flows \cite{Biferale2012,Chang2006,Shan1997}. 

In recent years, phase-field models are introduced as an effective modeling technique for interfacial physics. It uses an order parameter to distinguish different phases and postulates that the interface has finite width and material properties transit across the interfacial region smoothly but sharply. Traditional interface-capturing and interface-tracking methods are designed based on geometrical information of existing interfaces. When dealing with problems with phase transition, those methods become intractable. The phase-field models enjoy solid thermodynamic foundations \cite{Liu2014}, and this property allows them to describe complicated phase transition phenomena without resorting to ad hoc modeling tricks. The first instantiation of the phase-field models emanates from the work of van der Waals \cite{Waals1979} and Korteweg \cite{Korteweg1901}. It is now commonly known as the van der Waals fluid theory or the Navier-Stokes-Korteweg equations. In recent years, the van der Waals model was systematically analyzed in rational mechanics framework and the choice of the constitutive relations was carefully made to guarantee fundamental thermomechanical principles \cite{Dunn1985,Liu2013,Liu2015}. Recently progress has been made in boiling simulation using the Navier-Stokes-Korteweg equations. Boiling is regarded to be highly challenging for numerical simulations \cite{Juric1998,Lakkaraju2013}. It involves several physical mechanisms in multiple spatial-to-temporal scales. Most of the mechanisms are still not well understood quantitatively. Due to its improved approximation property for real fluids and its simplicity in describing boiling phenomena, it is appealing to investigate the free convection for the van der Waals fluids.

In this work, I will show that the van der Waals fluid model gives an accurate description of gas and liquid phases for various single-component fluids over a wide range of temperature. The satisfaction of the Clausius-Clapeyron relation demonstrates the model's capability in describing phase transition phenomena. The van der Waals model is incorporated into the continuum mechanics equations, and the resulting governing equations constitute a set of partial differential equations, which involve a third-order differential operator. The solid thermodynamic foundation allows this set of partial differential equations to describe bubble generation, phase transition, and topology change of interfaces in a unified approach. The numerical method for solving these equations is based on a provably unconditionally stable, second-order accurate scheme \cite{Liu2015}. Isogeometric analysis \cite{Hughes2005} is invoked to provide high-resolution spatial discretization. With the numerical method, I will first study the solutions of the Navier-Stokes-Korteweg equations at a small Rayleigh number. The steady state solution of the system at different volumetric averaged densities are solved and discussed. It will be shown that large volumetric averaged densities lead to stratified single-phase fluid as the steady state solution. In contrast, liquid-vapor two-phase fluid can be generated at low volumetric averaged densities. Next, two numerical simulations are performed to study the capability of the model in describing nucleate and film boiling. It will be shown that low fluid viscosity results in nucleate boiling while high fluid viscosity leads to film boiling. In the last, a suite of eighteen numerical simulations is performed to study the thermal convection property in the van der Waals fluid at volumetric averaged density $0.8$.

\section{Model}
\subsection{The van der Waals model}
In this work, a fixed, bounded, and connected domain $\Omega \subset \mathbb R^{n_d}$ is considered, where $n_d$ represents the number of space dimensions. The time interval of interest is denoted as $(0,T)$, with $T>0$. The Navier-Stokes-Korteweg equations are posed in the space-time domain $\Omega \times (0,T)$ as
\begin{eqnarray}
\label{eq:NSK_mass_divergence_form}
& & \frac{\partial \rho}{\partial t} + \nabla \cdot \left( \rho \mathbf u \right) = 0, \\
\label{eq:NSK_momentum_divergence_form}
& & \frac{\partial(\rho \mathbf u)}{\partial t}  + \nabla \cdot (\rho \mathbf u \otimes \mathbf u ) + \nabla p - \nabla \cdot  \bm \tau  - \nabla \cdot \bm \varsigma = \rho\mathbf g , \\
\label{eq:NSK_energy_divergence_form}
& & \frac{\partial (\rho E)}{\partial t} + \nabla \cdot \left( (\rho E + p) \mathbf{u}  - (\bm \tau + \bm \varsigma)\mathbf{u} \right) + \nabla \cdot \mathbf{q} + \nabla \cdot \bm \Pi  = \rho \mathbf{g} \cdot \mathbf{u} + \rho r.
\end{eqnarray}
The equations (\ref{eq:NSK_mass_divergence_form})-(\ref{eq:NSK_energy_divergence_form}) describe the balance of mass, linear momentum, and energy respectively. In these equations, $\rho$ is the fluid density, $\mathbf u$ is the velocity, $p$ is the thermodynamic pressure, $\bm \tau$ is the viscous stress, $\bm \varsigma$ is the Korteweg stress, $\mathbf g$ is the gravity, $E$ is the total energy per unit mass, $\mathbf q$ is the heat flux, and $\boldsymbol \Pi$ represents the power expenditure due to phase transitions \cite{Gurtin2009,Liu2015}. The constitutive relations for the van der Waals fluid are given as follows.
\begin{eqnarray}
\bm \tau &=& \bar{\mu} \left( \nabla \mathbf u + \nabla \mathbf u^{T} - \frac{2}{3}\nabla \cdot \mathbf u \mathbf I \right),   \\
\bm \varsigma &=& \left( \lambda \rho \Delta \rho + \frac{\lambda}{2}|\nabla\rho|^2 \right)\mathbf I - \lambda \nabla \rho \otimes \nabla \rho,  \\
\label{eq:def_pressure_constitutive}
p &=& Rb\theta \frac{\rho}{b-\rho} - a\rho^2,  \\
\mathbf q &=& - \kappa \nabla \theta, \\
\bm \Pi &=& \lambda \rho \nabla \cdot \mathbf u \nabla \rho,  \\
E &=& \iota + \frac{1}{2} |\mathbf u|^2,  \\
\iota &=& \iota_{loc} + \frac{\lambda}{2\rho}|\nabla \rho|^2,  \\
\label{eq:def_dim_iota_loc}
\iota_{loc} &=& -a\rho + C_v \theta.
\end{eqnarray}
In the above, $\bar{\mu}$ is the dynamic viscosity,  $\lambda$ is the capillarity coefficient, $R$ is the specific gas constant, $a$ and $b$ are fundamental fluid properties whose values for typical fluids can be found in \cite{Johnston2014}, $\kappa$ is the thermal conductivity, $C_v$ is the specific heat capacity at constant volume, $\iota$ is the internal energy per unit mass, and $\iota_{loc}$ is the local part of the internal energy density. 

Before proceeding further, let us introduce six additional thermodynamic quantities. First, the isobaric thermal expansion coefficient $\beta$ is defined as
\begin{eqnarray}
\beta := -\frac{1}{\rho} \frac{\partial \rho}{\partial \theta}.
\end{eqnarray}
Taking partial derivative with respect to $\theta$ at both sides of \eqref{eq:def_pressure_constitutive}, one can readily obtain an explicit expression of $\beta$ for the van der Waals fluid as
\begin{eqnarray}
\label{eq:vdw_thermal_expansion_coeff_def}
\beta = \frac{Rb(b-\rho)}{Rb^2\theta - 2a\rho(b-\rho)^2}.
\end{eqnarray} 
Second, the heat capacity at constant pressure is defined as
\begin{eqnarray}
C_p := \frac{\partial \iota_{loc}}{\partial \theta} - \frac{p}{\rho^2}\frac{\partial \rho}{\partial \theta}.
\end{eqnarray}
Invoking the constitutive relations \eqref{eq:def_pressure_constitutive} and \eqref{eq:def_dim_iota_loc}, one has
\begin{eqnarray}
\label{eq:vdw_heat_capacity_relation}
C_p := C_v  + \frac{Rb\theta}{b-\rho}\beta.
\end{eqnarray}
It is known that $C_v$ can be expressed as
\begin{eqnarray}
C_v = \varpi R,
\end{eqnarray}
wherein the non-dimensional number $\varpi$ depends on the structure of the fluid molecules \cite{Landau1987}. Hence one may further rewrite $C_p$ as 
\begin{eqnarray}
C_p &=& (\varpi + \frac{b\theta \beta}{b-\rho}) R = (\varpi + \chi) R, \\
\chi &:=& \frac{b \theta \beta}{b - \rho} = \frac{Rb^2\theta}{Rb^2\theta - 2a\rho(b-\rho)^2}.
\end{eqnarray}
Third, the local electrochemical potential is defined as
\begin{eqnarray}
\nu_{loc} := -2a\rho + R\theta
\log\left(\frac{\rho}{b-\rho} \right) -
C_v\theta\left( \log\left( \frac{\theta}{\theta_{\textup{ref}}} \right) -1 \right) + \frac{Rb\theta}{b-\rho},
\end{eqnarray}
wherein $\theta_{\textup{ref}}>0$ is the reference temperature. Fourth, the entropy for the van der Waals fluid is defined as
\begin{eqnarray}
s := -R \log\left(\frac{\rho}{b-\rho} \right) + C_v \log\left( \frac{\theta}{\theta_{\textup{ref}}} \right).
\end{eqnarray}
In the last, the kinematic viscosity $\bar{\nu}$ is defined as $\bar{\nu} := \bar{\mu} / \rho$. The thermal diffusivity $\alpha$ is defined as $\alpha := \kappa / C_p \rho$.
\begin{remark}
In the limit of $\rho \rightarrow 0$, the relations  \eqref{eq:vdw_thermal_expansion_coeff_def} and \eqref{eq:vdw_heat_capacity_relation} lead to
\begin{eqnarray*}
\beta \rightarrow \frac{1}{\theta}, \qquad C_p \rightarrow C_v + R.
\end{eqnarray*}
These two relations recover the thermal expansion coefficient and the Mayer's relation for ideal gases \cite{Landau1987}.
\end{remark}

\begin{figure}
	\begin{center}
	\begin{tabular}{c}
\includegraphics[angle=0, trim=130 60 130 120, clip=true, scale = 0.2]{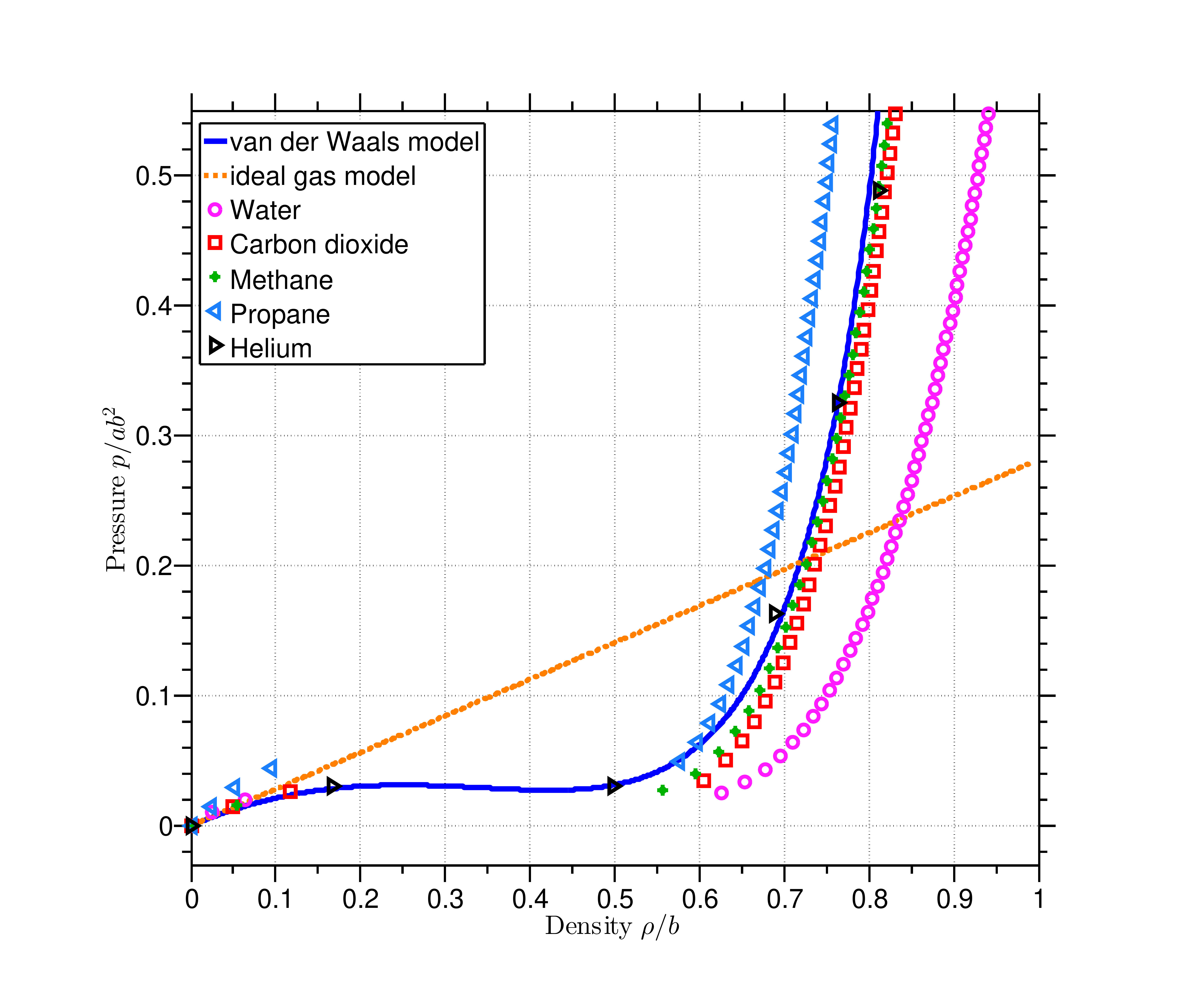} \\
(a)\\
\includegraphics[angle=0, trim=130 60 130 120, clip=true, scale = 0.2]{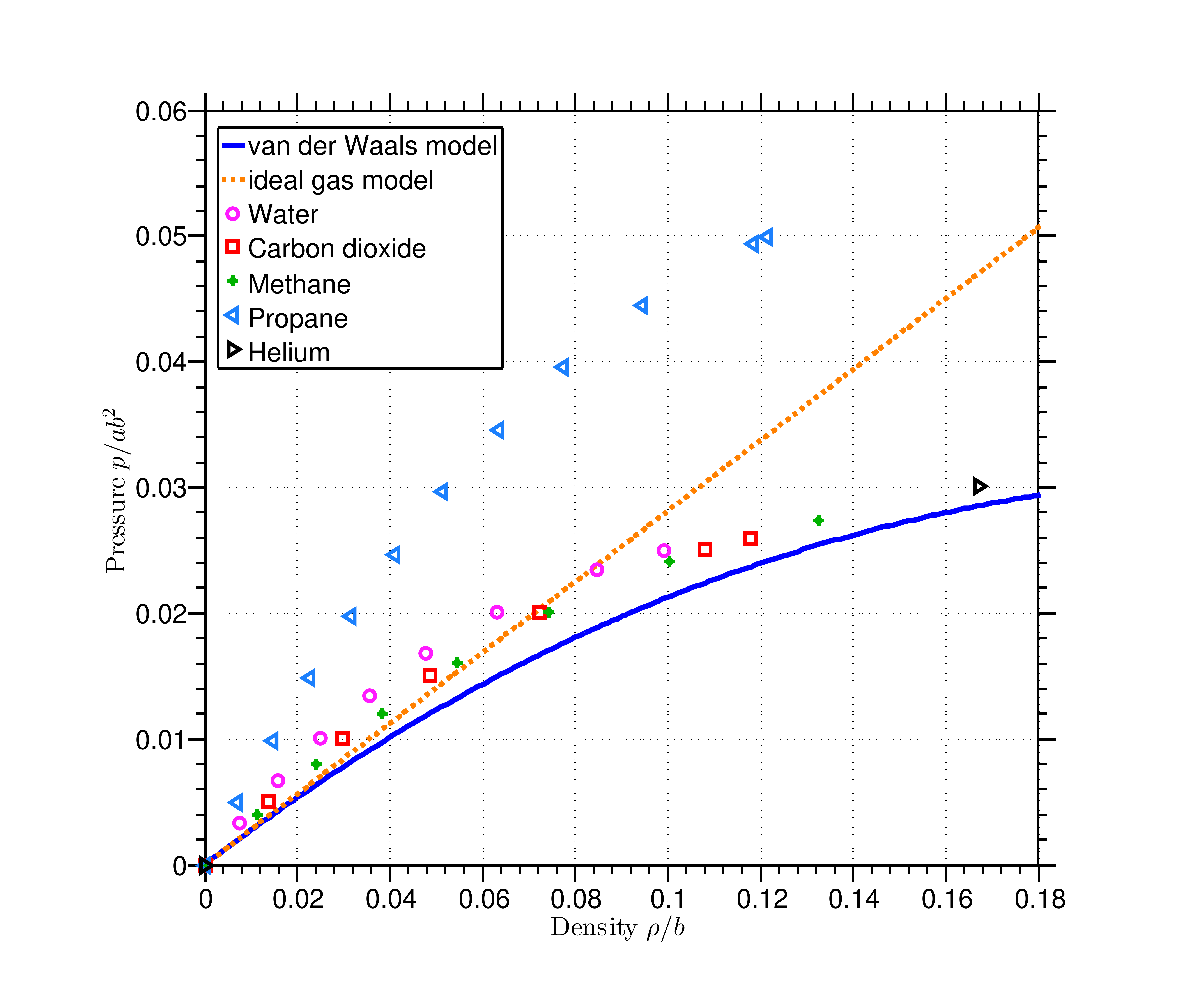}\\
(b)
\end{tabular}
\caption{Comparison of the van der Waals model and the ideal gas model  with real fluids at temperature $\theta = 0.95\theta_0$, $\theta_0 = 8ab / 27R$. The data for water, carbon dioxide, methane, propane, and helium are obtained from \cite{E.W.Lemmon2016} and scaled to dimensionless form. Fig. (b) gives a detailed view in the vapor phase.} 
\label{fig:vdw_compare_real_fluid_data_0d95}
\end{center}
\end{figure}

The van der Waals fluid model is regarded as a good approximation for fluids in both vapor and liquid phases. In Fig. \ref{fig:vdw_compare_real_fluid_data_0d95}, the van der Waals equation of state \eqref{eq:def_pressure_constitutive} is plotted as a uni-variate function of density at 0.95 of the critical temperature. The ideal gas model and real fluid data \cite{E.W.Lemmon2016} are plotted in the same figure for comparison purpose. It can be observed that the val der Waals model gives a very accurate description for various different fluids in both vapor and liquid phases. It is worth mentioning that the approximation is particularly excellent for helium, which is widely used as the working fluid in Rayleigh-B\'enard experiments \cite{Behringer1985}. The good approximation attribute makes the van der Waals model an ideal candidate for studying the Rayleigh-B\'enard convection for liquid-vapor two-phase fluids.

\subsection{Dimensional analysis}
\begin{figure}
	\begin{center}
	\begin{tabular}{c}
\includegraphics[angle=0, trim=130 90 130 120, clip=true, scale = 0.2]{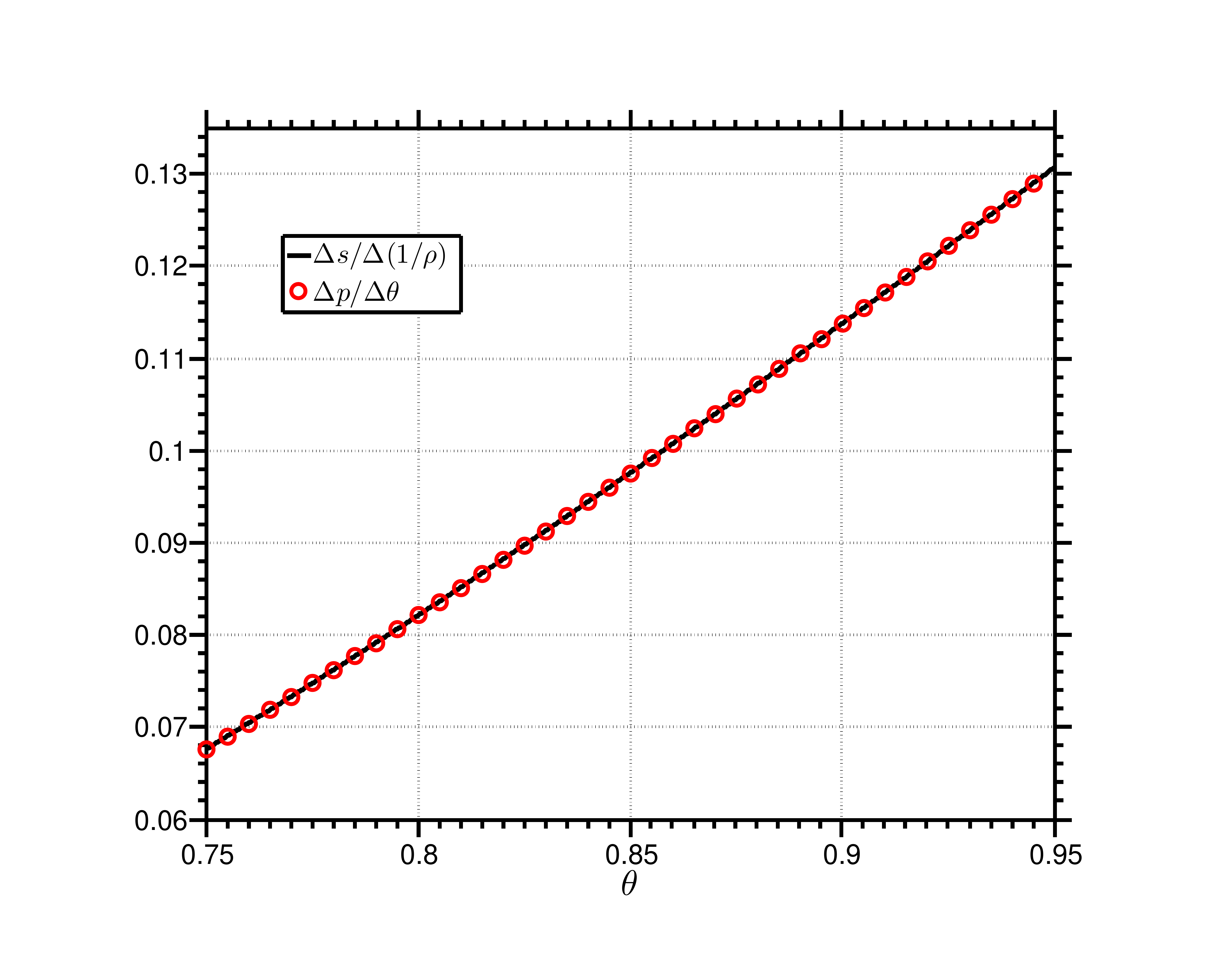} 
\end{tabular}
\caption{Clausius-Clapeyron relation at different temperatures. At a fixed temperature $\theta_0$, the liquid and vapor densities at the Maxwell state are denoted as $\rho^{\theta_0}_l$ and $\rho^{\theta_0}_v$. $\Delta s = s(\rho^{\theta_0}_v, \theta_0) - s(\rho^{\theta_0}_l, \theta_0)$, $\Delta(1/\rho) = 1/\rho^{\theta_0}_v - 1/\rho^{\theta_0}_l$. Let $p^{equ}(\theta)$ denote the equilibrium pressure at temperature $\theta$, $\Delta p / \Delta \theta$ at $\theta_0$ is evaluated using the central difference scheme: $p^{equ}(\theta_0+\Delta \theta) - p^{equ}(\theta_0 - \Delta \theta)/(2\Delta \theta)$.} 
\label{fig:clausius_clapeyron}
\end{center}
\end{figure}

In this section, dimensional analysis is performed for the system of equations \eqref{eq:NSK_mass_divergence_form}-\eqref{eq:NSK_energy_divergence_form} using $M_0$, $L_0$, $T_0$, and $\theta_0$ as the reference scales of mass, length, time, and temperature. If the reference scales are chosen as
\begin{eqnarray}
\label{eq:ref_scale_mtth}
\frac{M_0}{L_0^3} = b, \quad \frac{M_0}{L_0 T_0^2} = ab^2, \quad \theta_0 = \frac{8ab}{27R},
\end{eqnarray}
and $\theta_{\textup{ref}}$ is selected as $\theta_{\textup{ref}} = \theta_0$,
the dimensionless system can be written as
\begin{eqnarray}
\label{eq:dimless_nsk_mass}
& & \frac{\partial \rho^*}{\partial t^*} + \nabla^* \cdot \left( \rho^* \mathbf{u}^* \right) = 0, \\
\label{eq:dimless_nsk_momentum}
& & \frac{\partial(\rho^* \mathbf{u}^*)}{\partial t^*}  + \nabla^* \cdot (\rho^* \mathbf{u}^* \otimes \mathbf{u}^* ) + \nabla^* p^* - \nabla^* \cdot \boldsymbol \tau^* - \nabla^* \cdot \boldsymbol \varsigma^* = \rho^*\mathbf{g}^*, \\
\label{eq:dimless_nsk_energy}
& & \frac{\partial (\rho^* E^*)}{\partial t^*} + \nabla^* \cdot \left( (\rho^* E^* + p^*) \mathbf{u}^*  - (\boldsymbol \tau^* + \boldsymbol \varsigma^*)\mathbf{u}^* \right) + \nabla^* \cdot \mathbf{q}^* + \nabla^* \cdot \boldsymbol \Pi^*  \nonumber \\
& & \hspace{3cm} = \rho^* \mathbf{g}^* \cdot \mathbf{u}^* + \rho^* r^*,
\end{eqnarray}
wherein,
\begin{align}
p^* &= \frac{8\theta^*\rho^*}{27(1-\rho^*)} - \rho^{*2}, \\
\boldsymbol \tau^* &= \bar{\mu}^* \left( \nabla^* \mathbf u^* + \nabla^* \mathbf u^{*T} - \frac{2}{3}\nabla^* \cdot \mathbf u^* \mathbf I \right), \\
\boldsymbol \varsigma^* &= \lambda^* \left( \left( \rho^* \Delta^* \rho^* + \frac{1}{2}|\nabla^* \rho^*|^2 \right)\mathbf I - \nabla^* \rho^* \otimes \nabla^* \rho^* \right), \\
\mathbf q^* &= - \kappa^* \nabla^* \theta^*, \displaybreak[3] \\
\label{eq:tnsk_dimenless_form_interstitial_term_with_star}
\bm \Pi^* &= \lambda^* \rho^* \nabla^* \cdot \mathbf u^*
 \nabla^* \rho^*, \\
\label{eq:dimless_def_iota_loc}
\iota^* &=  \iota^*_{loc} + \frac{\lambda^*}{2\rho^*}|\nabla^* \rho^*|^2, \\
\iota^*_{loc} &= -\rho^* + \frac{8\varpi}{27}\theta^*, \displaybreak[3] \\
\label{eq:def_nu_loc}
\nu^*_{loc} &= -2\rho^* + \frac{8\theta^*}{27(1-\rho^*)} + \frac{8}{27}\theta^*\log\left(\frac{\rho^*}{1-\rho^*} \right) + \frac{8\varpi}{27}\theta^* \left( 1 - \log(\theta^*)\right), \displaybreak[3] \\
s^* &= -\frac{8}{27}\log \left(\frac{\rho^*}{1-\rho^*} \right) + \frac{8\varpi}{27}\log \left(\theta^* \right), \displaybreak[3] \\
\bar{\mu}^* &= \frac{\bar{\mu}}{L_0 b \sqrt{ab}}, \displaybreak[3] \\
\lambda^* &= \frac{\lambda}{a L_0^2},  \displaybreak[3] \\
\kappa^* &= \kappa \frac{8}{27R (ab)^{1/2}bL_0},  \displaybreak[3] \\
\mathbf g^* &= \mathbf g \frac{L_0}{ab}.
\end{align}
The dimensionless isobaric thermal expansion coefficient $\beta$ can be expressed as
\begin{eqnarray}
\beta &=& \frac{c_{\beta}}{\theta_0}, \\
c_{\beta} &=& c_{\beta}(\rho^*,\theta^*) = \frac{4(1-\rho^*)}{4\theta^* -27\rho^*(1-\rho^*)^2}.
\end{eqnarray}
The heat capacity at constant pressure can be expressed as
\begin{eqnarray}
C_p &=& C_v + \chi R = (\varpi + \chi) R, \\
\chi &=& \chi(\rho^*,\theta^*) = \frac{c_{\beta}\theta^*}{1-\rho^*} = \frac{4\theta^*}{4\theta^* -27\rho^*(1-\rho^*)^2}.
\end{eqnarray}
The non-dimensional kinematic viscosity and thermal diffusivity are given as
\begin{eqnarray}
\bar{\nu}^* = \frac{\bar{\nu}}{L_0 \sqrt{ab}}, \qquad \alpha^* = \alpha \frac{8\left(\varpi + \chi \right)}{27\sqrt{ab}L_0}.
\end{eqnarray}
At a given temperature, the equilibrium states of vapor and liquid can be found by equating the pressure and chemical potential \cite{Johnston2014,Liu2015}. This solution procedure involves solving a system of two nonlinear algebraic equations. Table \ref{table:tnsk_maxwell_states_properties} shows the coexistent vapor and liquid densities and the corresponding values of $c_{\beta}$ and $\chi$ at several different temperatures. For multiphase fluids, the Clausius-Clapeyron relation characterizes the energy released or absorbed during a phase transition process. It relates the latent heat $\Delta s \theta_0$ with the coexistence curve in the pressure-temperature diagram. In Fig. \ref{fig:clausius_clapeyron}, the Clausius-Clapeyron relation for the van der Waals fluid at various temperature is illustrated, demonstrating the capability of the van der Waals fluid in modeling phase transition phenomena.

\begin{table}[htbp]
\begin{center}
\tabcolsep=0.19cm
\renewcommand{\arraystretch}{1.5}
\begin{tabular}{c c c c c c c}
\hline
\hline
$\theta^*$ & $\rho_v^*$ & $\rho_l^*$ & $c_{\beta}(\rho_v^*, \theta^*)$ & $c_{\beta}(\rho_l^*, \theta^*)$ & $\chi(\rho_v^*, \theta^*)$  & $\chi(\rho_l^*, \theta^*)$ \\
\hline
0.990 & 0.2682 & 0.4012 & 35.6607 &  31.5997 & 48.2414 & 52.2409 \\
0.970 & 0.2228 & 0.4519 & 12.6242 & 10.2064 &  15.7562 & 18.0642 \\
0.950 & 0.1930 & 0.4872 & 7.9449 & 6.0123 &  9.3528 &  11.1391 \\
0.900 & 0.1419 & 0.5524 & 4.4078 & 2.9250 &  4.6231 & 5.8817 \\
0.865 & 0.1161 & 0.5884 &  3.4974 & 2.1422 & 3.4226 & 4.5020 \\
0.850 & 0.1066 & 0.6024 & 3.2397 & 1.9195 &  3.0822 & 4.1033 \\
0.800 & 0.0799 & 0.6442 & 2.6789 & 1.4253 &  2.3292 & 3.2051 \\
0.750 & 0.0591 & 0.6808 & 2.3701 & 1.1330 &  1.8892 & 2.6620 \\
0.700 & 0.0427 & 0.7135 & 2.1957 & 0.9405 &  1.6055 & 2.2978 \\
\hline 
\hline
\end{tabular}
\end{center}
\caption{The vapor and liquid densities at the Maxwell states are given for various temperatures. The values of the corresponding $C_{\beta}$ and $\chi$ are also evaluated.}
\label{table:tnsk_maxwell_states_properties}
\end{table}

In this study, the reference length scale $L_0$ is chosen such that $|\mathbf g^*|$ is $-0.025$, wherein $|\cdot|$ denotes the $l^2$-norm of a vector. The rest three reference scales can be determined correspondingly from \eqref{eq:ref_scale_mtth}. Hence, the evolution of the system \eqref{eq:dimless_nsk_mass}-\eqref{eq:dimless_nsk_energy} is governed by four non-dimensional parameters $\bar{\mu}^*$, $\lambda^*$, $\varpi$, and $\kappa^*$ (or equivalently, $\bar{\nu}^*$, $\lambda^*$, $\varpi$, and $\alpha^*$), the initial and boundary conditions, and the geometry of the domain. In this work, I consider the fluid dynamics constrained by two parallel plates. The gravity points in the direction opposite to the $z$-axis. The height between the two plates is denoted as $H = H^* L_0$; the length of the plates is denoted $D = D^* L_0$. The temperature on the bottom and top plates is denoted as $\theta_{b}$ and $\theta_{t}$ respectively, and the temperature difference is $\Delta \theta := \theta_b - \theta_t = \theta_0 (\theta_b^* - \theta_t^*) = \theta_0 \Delta \theta^*$. The aspect ratio $\Lambda$ characterizing the geometry of the domain is defined as
\begin{eqnarray}
\Lambda := \frac{D}{H} = \frac{D^*}{H^*}.
\end{eqnarray}
The Rayleigh number is defined as
\begin{eqnarray}
\operatorname{Ra} := \frac{g^* \Delta \theta^*}{\kappa^* \bar{\mu}^*} = \frac{g^* \Delta \theta^*}{\rho^{*2}\alpha^* \bar{\nu}^*}.
\end{eqnarray}
The Rayleigh number measures the relative strength of the buoyancy force in comparison with the resistant effect due to the viscosity and conductivity. The Prandtl number in the system is defined as
\begin{eqnarray}
\operatorname{Pr} := \frac{\bar{\mu}^*}{\kappa^*} = \frac{\bar{\nu}^*}{\alpha^*},
\end{eqnarray}
which measures the ratio of the momentum diffusivity to the thermal conductivity. In addition to the temperature difference $\Delta \theta^*$, the values of the temperature on the top and bottom boundaries will also affect the dynamics of the system since these values may change the liquid-vapor coexistence state (see Table \ref{table:tnsk_maxwell_states_properties}). Hence, I introduce the arithmetic mean of the top and bottom temperature as one control parameter:
\begin{eqnarray}
\theta_m := \frac{\theta_b + \theta_t}{2}.
\end{eqnarray}
The initial state of the total mass within the box is another control parameter for the system. It is described by the volumetric averaged density:
\begin{eqnarray}
\rho_m := \frac{\int_{\Omega}\rho d\mathbf x}{\int_{\Omega}d\mathbf x}.
\end{eqnarray}
A major question people strive to answer is the dependence of the heat transport efficiency and the flow structure on the control parameters in the van der Waals fluid. The heat transport efficiency is described by the Nusselt number, which measures the relative strength of the total heat flux in comparison with the purely diffusive heat flux. The heat flux can be measured either over a fixed horizontal plane or in the whole body. The Nusselt number measured at a horizontal plane $A$ is defined as
\begin{eqnarray}
\operatorname{Nu}_{A,t} := \frac{\left\langle  u_z \theta \right\rangle_{A,t} - \alpha \partial_z \left\langle \theta\right\rangle_{A,t} }{\alpha \Delta \theta H^{-1}} = \frac{ 8(\varpi + \chi)\rho^* \left\langle u_z^* \theta^* \right\rangle_{A,t} - 27 \kappa^* \partial_{z^*}\left\langle \theta^* \right\rangle_{A,t} }{27\kappa^* \Delta \theta^* H^{*-1}}.
\end{eqnarray}
Here, $u_z$ is the velocity component in the $z$-direction; $\left\langle  \cdot \right\rangle_{A,t}$ denotes an average operator over a horizontal plane $A$ and over time for a quantity. If one average $\operatorname{Nu}_{A,t}$ in the $z$-direction, the volume-averaged Nusselt number is obtained as
\begin{eqnarray}
\operatorname{Nu}_{V,t} := \frac{ 8(\varpi + \chi)\rho^* \left\langle u_z^* \theta^* \right\rangle_{V,t} }{27\kappa^* \Delta \theta^* H^{*-1}} + 1.
\end{eqnarray}
Here, $\left\langle  \cdot \right\rangle_{V,t}$ represents an average operator over the volume and over time. Henceforth, I will restrict my discussion to the dimensionless form, and the superscript $*$ will be omitted for notational simplicity.


\section{Numerical methods}
The numerical method for solving the governing equations \eqref{eq:dimless_nsk_mass}-\eqref{eq:dimless_nsk_energy} is based on a set of functional entropy variables. The mathematical entropy function associated with the van der Walls fluid is 
\begin{eqnarray*}
H := -\rho s = \frac{8}{27}\rho\log\left(\frac{\rho}{1-\rho} \right) - \frac{8\varpi}{27}\rho\log\left(\theta \right).
\end{eqnarray*}
For three-dimensional problems, the vector of conservation variables is
\begin{eqnarray*}
\bm U^T = [U_1, U_2, U_3, U_4, U_5] := [\rho,  \rho u_1, \rho u_2, \rho u_3 , \rho E].
\end{eqnarray*}
The entropy variables $\bm V^T= [V_1, V_2, V_3, V_4, V_5]$ are defined as the functional derivatives of $H$ with respect to $\bm U$:
\begin{eqnarray*}
V_i[\delta v_i] = \frac{\delta H}{\delta U_i}[\delta v_i], \quad i=1,\dots, 5,
\end{eqnarray*}
wherein $\delta \bm v^T = [\delta v_1, \delta v_2, \delta v_3, \delta v_4, \delta v_5 ]$ are the test functions. The entropy variables $\bm V$ can be written explicitly as
\begin{eqnarray*}
V_1[\delta v_1] &=& \frac{1}{\theta}\left( \nu_{loc} - \frac{|\bm u|^2}{2} \right) \delta v_1 + \lambda \frac{1}{\theta} \nabla
\rho \cdot \nabla \delta v_1,  \\
V_i[\delta v_i] &=& \frac{u_{i-1}}{\theta} \delta v_i, \quad i = 2, 3, 4,  \quad
V_5[\delta v_5] = -\frac{1}{\theta} \delta v_5.
\end{eqnarray*}
The definition of the entropy variable $V_1$ involves a non-local differential operator. Inspired from this fact, a new independent variable $V$ is introduced as
\begin{eqnarray*}
V := \frac{1}{\theta} \left( \nu_{loc} - \frac{|\mathbf u|^2}{2} \right) -
\lambda \nabla \cdot \left( \frac{\nabla \rho}{\theta} \right).
\end{eqnarray*} 
The fundamental thermodynamic relation between $p$ and $\nu_{loc}$ allows us to express $p$ in terms of $V$ as
\begin{equation}
\label{eq:tnsk_p_v_relation}
p = \rho V\theta - \rho \Psi_{loc} + \frac{\rho|\mathbf u|^2}{2}
+\lambda  \rho \theta \nabla \cdot \left( \frac{\nabla
\rho}{\theta} \right).
\end{equation}
Making use of the relation (\ref{eq:tnsk_p_v_relation}), the original strong-form problem (\ref{eq:dimless_nsk_mass})-(\ref{eq:dimless_nsk_energy}) can be rewritten as
\begin{eqnarray}
\label{eq:mtnsk_mass}
& & \frac{\partial \rho}{\partial t} + \nabla \cdot (\rho \mathbf u) = 0, \\
\label{eq:mtnsk_momentum}
& & \frac{\partial (\rho \mathbf u)}{\partial t} + \nabla \cdot \left( \rho \mathbf u \otimes \mathbf u \right) + \nabla \left( \rho V \theta + \frac{\rho |\mathbf u|^2}{2} + \lambda \rho \theta \nabla \cdot \left( \frac{\nabla\rho}{\theta} \right) \right) \nonumber \\
& & \hspace{5mm} - \left( V\theta + \frac{|\mathbf u|^2}{2} +
\lambda \theta \nabla \cdot \left( \frac{\nabla \rho}{\theta}
\right) \right)\nabla \rho -  H \nabla \theta - \nabla \cdot
\bm \tau - \nabla \cdot \bm \varsigma = \rho \mathbf b,  \\
\label{eq:mtnsk_energy}
& & \frac{\partial (\rho E)}{\partial t} + \nabla \cdot \left( \left(\rho V \theta - \theta H  + \lambda |\nabla \rho|^2  + \rho|\mathbf u|^2 + \lambda \rho \theta \nabla \cdot \left( \frac{\nabla \rho}{\theta} \right) \right) \mathbf u \right) \nonumber   \\
& & \hspace{5mm} - \nabla \cdot \left( \left( \bm \tau + \bm \varsigma \right) \mathbf u \right) + \nabla \cdot \mathbf q + \nabla \cdot \boldsymbol \Pi = \rho \mathbf b \cdot \mathbf u + \rho r,   \\
\label{eq:mtnsk_aux}
& & V = \frac{1}{\theta} \left( \nu_{loc} - \frac{|\mathbf u|^2}{2} \right) - \lambda \nabla \cdot \left( \frac{\nabla \rho}{\theta} \right).
\end{eqnarray}
The new strong-form problem \eqref{eq:mtnsk_mass}-\eqref{eq:mtnsk_aux} is an equivalent statement of the original Navier-Stokes-Korteweg equations \eqref{eq:dimless_nsk_mass}-\eqref{eq:dimless_nsk_energy}. Based on this new strong-form problem, the numerical scheme can be constructed. Let the time interval $(0,T)$ be divided into $N_{ts}$ subintervals  $(t_n, t_{n+1})$, $n=0, \cdots, N_{ts}-1$, of size $\Delta t_n = t_{n+1} -t_n$.  I use the notation
\begin{eqnarray*}
\bm Y^h_n := \left[ \rho^h_n, \frac{u^h_{1,n}}{\theta^h_n} , \frac{u^h_{2,n}}{\theta^h_n} , \frac{u^h_{3,n}}{\theta^h_n}, \frac{-1}{\theta^h_n} , V^h_n \right]^T
\end{eqnarray*}
to represent the fully discrete solutions at the time level $n$. I define the jump of density, linear momentum, and total energy over each time step as
\begin{eqnarray*}
\llbracket \rho^h_n \rrbracket &&:= \rho^h_{n+1} - \rho^h_{n}, \quad
\llbracket \rho^h_n \mathbf u^h_n \rrbracket := \rho^h_{n+1} \mathbf u^h_{n+1} - \rho^h_{n} \mathbf u^h_{n}, \\
\big[\rho^h_n E(\rho^h_n, \mathbf u^h_n, \theta^h_n)\big]  &&:= (\rho\Psi_{loc})(\rho^h_{n+\frac{1}{2}},\theta^h_{n+1}) - (\rho\Psi_{loc})(\rho^h_{n+\frac{1}{2}},\theta^h_{n}) \nonumber \\
& & + (\rho\Psi_{loc})(\rho^h_{n+1},\theta^h_{n+\frac{1}{2}}) - (\rho\Psi_{loc})(\rho^h_{n},\theta^h_{n+\frac{1}{2}}) \nonumber \\
& & - \theta^h_{n+\frac{1}{2}} \left( H(\rho^h_{n+1}, \theta^h_{n+1}) - H(\rho^h_{n}, \theta^h_{n}) \right) \nonumber \\
& & - \frac{\theta^h_{n+1}-\theta^h_n}{2} \left( H(\rho^h_{n+\frac{1}{2}},\theta^h_{n+1}) + H(\rho^h_{n+\frac{1}{2}},\theta^h_{n})  \right) \nonumber \\
& & + \frac{(\theta^h_{n+1} - \theta^h_n)^3}{12}\frac{\partial^2 H}{\partial \theta^2}(\rho^h_{n+\frac12}, \theta^h_{n+1}) \nonumber \\
& & + \frac{1}{2}\left( \rho^h_{n+1}|\mathbf u^h_{n+1}|^2 - \rho^h_n |\mathbf u^h_n|^2\right)  + \lambda \left( |\nabla \rho^h_{n+1}|^2 - |\nabla \rho^h_{n}|^2 \right).
\end{eqnarray*}
With the jump operators defined above, the fully discrete scheme can be stated as follows. In each time step, given $\mathbf Y^h_n$ and the time step $\Delta t_n$, find $\mathbf Y^h_{n+1}$ such that for all $w_1^h \in \mathcal V^h$, $\mathbf w^h=(w_2^h; w_3^h; w_4^h)^T \in \left( \mathcal V^h\right)^3$, $w_5^h \in \mathcal V^h$, and $w_6^h \in \mathcal V^h$,
\begin{align}
\label{eq:mtnsk_fully_discrete_mass}
& \mathbf B^M(w_1^h; \mathbf Y^h_{n+1}) := \left( w_1^h, \frac{\llbracket \rho^h_n \rrbracket }{\Delta t_n} \right)_{\Omega} - \left( \nabla w_1^h, \rho^h_{n+\frac12} \mathbf u^h_{n+\frac12} \right)_{\Omega} = 0, \displaybreak[2] \\
\label{eq:mtnsk_fully_discrete_momentum}
& \mathbf B^U(\mathbf w^h; \mathbf Y^h_{n+1}) := \left( \mathbf w^h, \frac{\llbracket \rho^h_n \mathbf u^h_n \rrbracket}{\Delta t_n} \right)_{\Omega} - \left( \nabla \mathbf w^h, \rho^h_{n+\frac12}\mathbf u^h_{n+\frac12} \otimes \mathbf u^h_{n+\frac12} \right)_{\Omega} \nonumber  \displaybreak[2] \\
&  - \left( \nabla \cdot \mathbf w^h, \rho^h_{n+\frac12} V^h_{n+\frac12}\theta^h_{n+\frac12} + \frac{1}{2}\rho^h_{n+\frac12} |\mathbf u^h_{n+\frac12}|^2 + \lambda \rho_{n+\frac12}^h \theta_{n+\frac12}^h \nabla \cdot \left(\frac{\nabla\rho_{n+\frac12}^h}{\theta_{n+\frac12}^h} \right) \right)_{\Omega} \nonumber   \displaybreak[2] \\
&  - \left( \mathbf w^h, \left( V_{n+\frac12}^h\theta_{n+\frac12}^h + \frac{|\mathbf u_{n+\frac12}^h|^2}{2} + \lambda \theta_{n+\frac12}^h \nabla \cdot \left(\frac{\nabla\rho_{n+\frac12}^h}{\theta_{n+\frac12}^h} \right)\right) \nabla \rho_{n+\frac12}^h \right)_{\Omega} \nonumber \displaybreak[2] \\
&  - \left( \mathbf w^h, H_{n+\frac12}^h \nabla \theta_{n+\frac12}^h \right)_{\Omega} + \left(\nabla \mathbf w^h, \boldsymbol \tau_{n+\frac12}^h + \boldsymbol \varsigma_{n+\frac12}^h \right)_{\Omega} - \left(\mathbf w^h, \rho_{n+\frac12}^h \mathbf b\right)_{\Omega} = \mathbf 0,  \displaybreak[2] \\
\label{eq:mtnsk_fully_discrete_energy}
&  \mathbf B^E(w_5^h; \mathbf Y^h_{n+1}) := \left( w_5^h, \frac{\big[ \rho^h_n E(\rho^h_n, \mathbf u^h_n, \theta^h_n)\big]}{\Delta t_n} \right)_{\Omega} - \Bigg( \nabla w^h_5, \Bigg( \rho_{n+\frac12}^h V_{n+\frac12}^h \theta_{n+\frac12}^h     \nonumber  \displaybreak[2] \\
&  - \theta_{n+\frac12}^h  H_{n+\frac12}^h  + \frac{\lambda |\nabla \rho_{n+\frac12}^h|^2}{2} +  \lambda \rho_{n+\frac12}^h \theta_{n+\frac12}^h  \nabla \cdot \left( \frac{\nabla \rho_{n+\frac12}^h}{\theta_{n+\frac12}^h} \right) + \rho_{n+\frac12}^h|\mathbf u_{n+\frac12}^h|^2 \Bigg) \mathbf u_{n+\frac12}^h \Bigg)_{\Omega} \nonumber  \displaybreak[2]  \\
&  + \left( \nabla w^h_5, \boldsymbol \tau_{n+\frac12}^h \mathbf u_{n+\frac12}^h \right)_{\Omega} + \left(\nabla w^h_5, \boldsymbol \varsigma_{n+\frac12}^h \mathbf u_{n+\frac12}^h \right)_{\Omega} - \left(\nabla w^h_5, \mathbf q_{n+\frac12}^h + \boldsymbol \Pi_{n+\frac12}^h \right)_{\Omega}  \nonumber  \displaybreak[2] \\
&  - \left(w^h_5, \rho_{n+\frac12}^h \mathbf b \cdot \mathbf u_{n+\frac12}^h \right)_{\Omega}  - \left(w^h_5, \rho_{n+\frac12}^h r\right)_{\Omega} = 0,  \displaybreak[2] \\
\label{eq:mtnsk_fully_discrete_aux}
& \mathbf B^A(w_6^h; \mathbf Y^h_{n+1}) := \Bigg( w^h_6, V_{n+\frac12}^h - \frac{1}{2\theta_{n+\frac12}^h} \bigg( \Big(\nu_{loc}(\rho^h_n, \theta^h_{n+\frac12}) + \nu_{loc}(\rho^h_{n+1}, \theta^h_{n+\frac12})  \Big)  \nonumber \displaybreak[2] \\
& + \frac{\llbracket \rho^h_n \rrbracket^2}{12}\frac{\partial^2\nu_{loc}}{\partial \rho^2}(\rho^h_n, \theta^h_{n+\frac12}) \bigg) - \frac{\mathbf u^h_n \cdot \mathbf u^h_{n+1}}{2\theta^h_{n+\frac12}} \Bigg)_{\Omega} - \left( \nabla w^h_6, \frac{ \lambda \nabla \rho^h_{n+\frac12}}{\theta^h_{n+\frac12}} \right)_{\Omega} = 0.
\end{align}
The main results of the fully discrete scheme (\ref{eq:mtnsk_fully_discrete_mass})-(\ref{eq:mtnsk_fully_discrete_aux}) are stated in the following two theorems.
\begin{theorem}
\label{theorem:tnsk_thermodynamic_consistency}
The solutions of the fully discrete scheme (\ref{eq:mtnsk_fully_discrete_mass})-(\ref{eq:mtnsk_fully_discrete_aux}) satisfy
\begin{eqnarray*}
\label{eq:tnsk_thermodynamic_consistency_theorem}
&&\int_{\Omega} \Bigg( \frac{ H(\rho^h_{n+1},\theta^h_{n+1}) -  H(\rho^h_n, \theta^h_n)}{\Delta t_n} + \nabla \cdot \left( H(\rho^h_{n+\frac12},\theta^h_{n+\frac12}) \mathbf u_{n+\frac12}^h \right) - \nabla \cdot \left(\frac{\mathbf q_{n+\frac12}^h}{\theta^h_{n+\frac12}} \right) \nonumber   \\
&&\qquad + \frac{\rho_{n+\frac12}^h r}{\theta^h_{n+\frac12}} \Bigg) dV_\mathbf x = -\int_{\Omega} \frac{1}{\theta^h_{n+\frac12}} \boldsymbol \tau_{n+\frac12}^h : \nabla \mathbf u^h_{n+\frac12} dV_\mathbf x  - \int_{\Omega} \frac{ \kappa |\nabla \theta_{n+\frac12}^h|^2}{\left(\theta_{n+\frac12}^{h}\right)^2}dV_\mathbf x \nonumber \\
&&- \int_{\Omega}\frac{1}{\theta^h_{n+\frac12}\Delta t_n} \left( \frac{\llbracket \rho^h_n \rrbracket^4}{24}\frac{\partial^3\nu_{loc}}{\partial \rho^3}(\rho^h_{n+\xi_1},\theta^h_{n+\frac12})  - \frac{\llbracket \theta^h_n \rrbracket^4}{24}\frac{\partial^3 H}{\partial \theta^3}(\rho^h_{n+\frac12},\theta^h_{n+{\xi_2}}) \right) dV_\mathbf x \\
&&  \leq 0.
\end{eqnarray*}
\end{theorem}

\begin{theorem}
\label{theorem:tnsk_time_accuracy}
The local truncation error in time $\bm \Theta(t) = \left( \Theta_\rho(t); \bm \Theta^T_{\mathbf u}(t); \Theta_E(t) \right)^T$ is bounded by $|\bm \Theta(t_n)| \leq K \Delta t_n^2\mathbf 1_5$ for all $t_n \in [0,T]$, where $K$ is a constant independent of $\Delta t_n$ and $\mathbf 1_5 = (1;1;1;1;1)^T$.
\end{theorem}
The above two theorems are proven in \cite{Liu2015}. Theorem 1 states that the numerical method is unconditionally stable in entropy. Theorem 2 states that the temporal scheme is second-order accurate. The numerical scheme is implemented based on the PETSc package \cite{Balay1997}, and code verification has been performed a thorough comparison with manufactured solutions and ``overkill" solutions \cite{Liu2015}. In this study, $C^1$-continuous quadratic B-splines are employed to define $\mathcal V^h$ as well as the computational domain. Consequently, this approach may be considered as the isogeometric analysis method \cite{Hughes2005}. In all simulations, I fix $\varpi = 3$, $\Lambda = 2$. On boundary surfaces, ninety-degree contact angle boundary condition $\nabla \rho \cdot \mathbf n =0$ is imposed for the density variable; no-slip boundary condition is imposed for the velocity; Dirichlet boundary condition is imposed for the temperature on the top and bottom surfaces, and the adiabatic condition is imposed on the vertical boundary surfaces. 

\begin{figure}[htp!]
\begin{center}
\begin{tabular}{cc}
\scalebox{0.33}{\includegraphics[angle=0, trim=60 470 60 70,
clip=true]{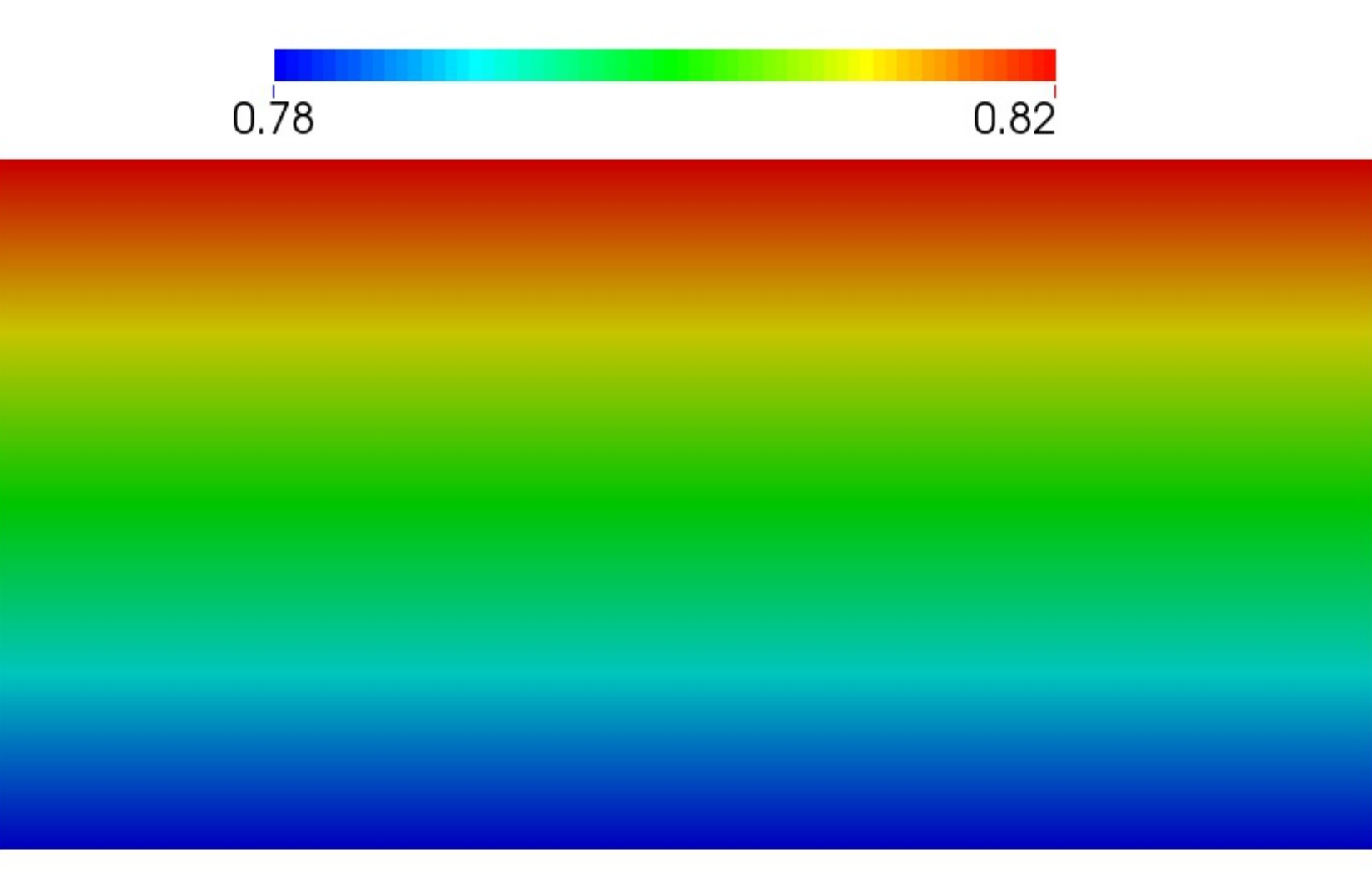} } & 
\scalebox{0.33}{\includegraphics[angle=0, trim=60 470 60 70,
clip=true]{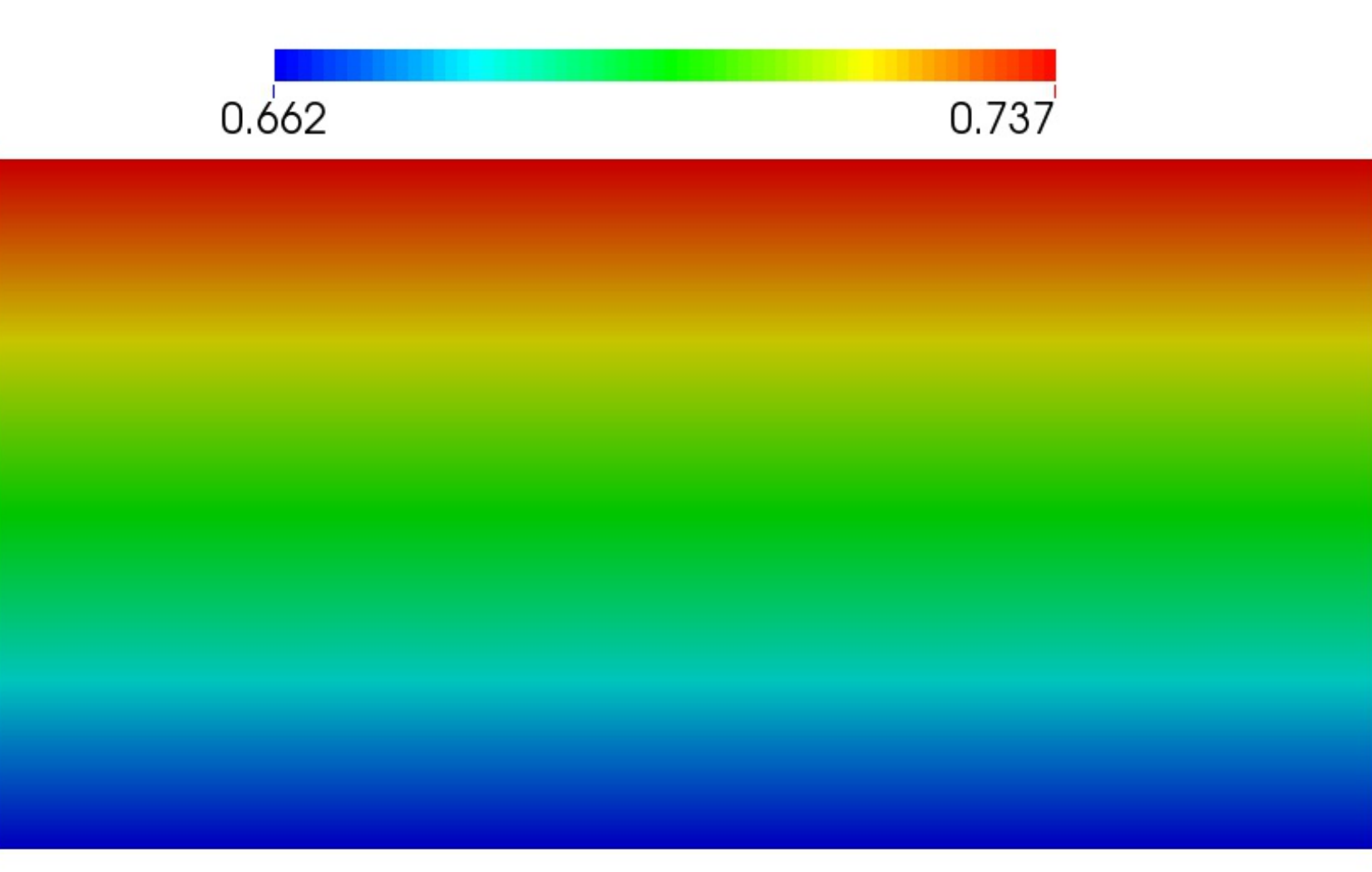} } \\
(a)   & (b) \\ 
\scalebox{0.33}{\includegraphics[angle=0, trim=60 470 60 70,
clip=true]{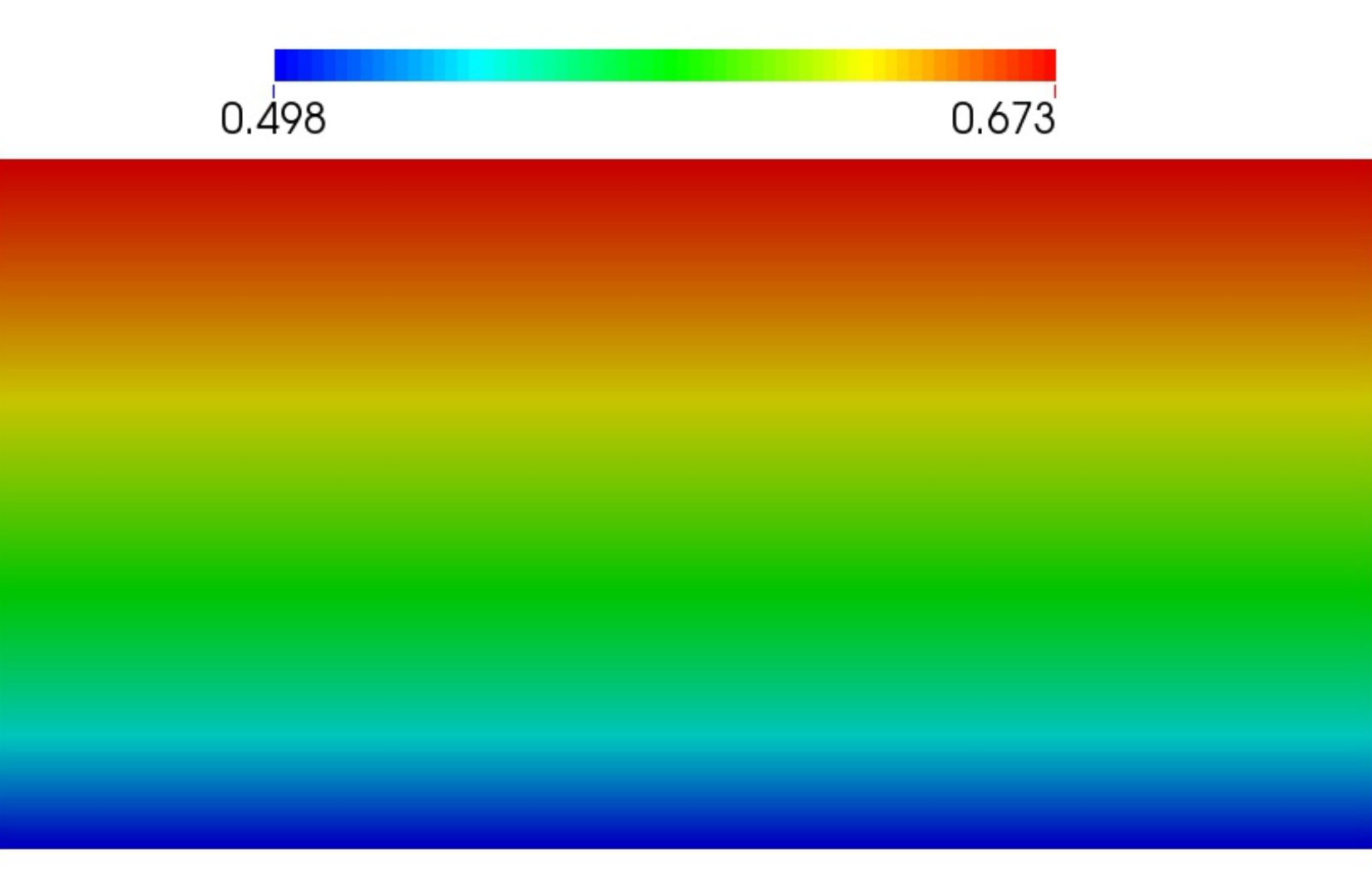} } & 
\scalebox{0.33}{\includegraphics[angle=0, trim=60 470 60 70,
clip=true]{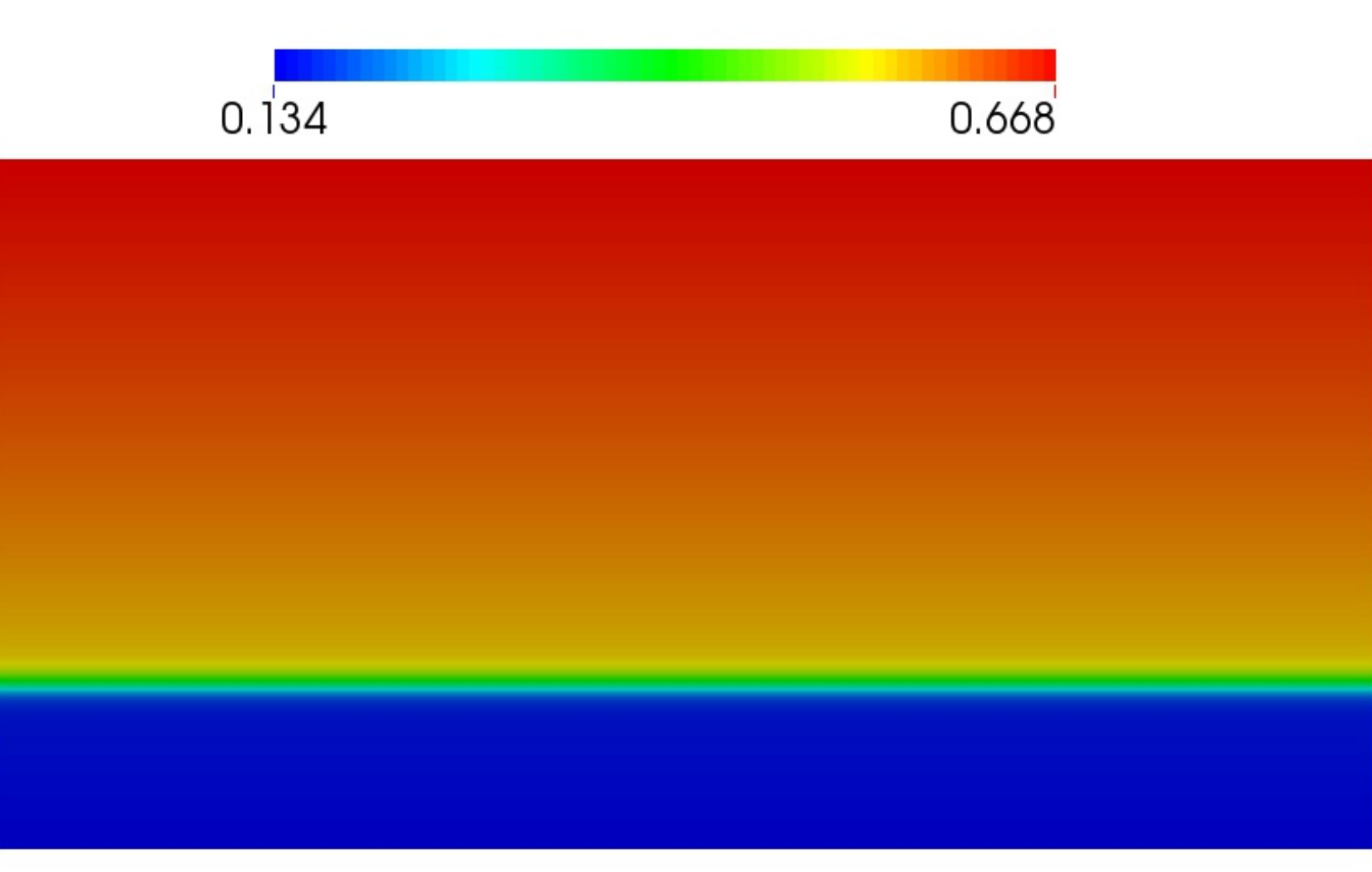} } \\
(c)   & (d)
\end{tabular}
\caption{The density profile of the steady state solution for (a) $\rho_m = 0.8$, (b) $\rho_m = 0.7$, (c) $\rho_m = 0.6$, and (d) $\rho_m = 0.5$. }
\label{fig:smallRa_all}
\end{center}
\end{figure}

\begin{figure}
	\begin{center}
	\begin{tabular}{c}
\includegraphics[angle=0, trim=110 125 130 160, clip=true, scale = 0.23]{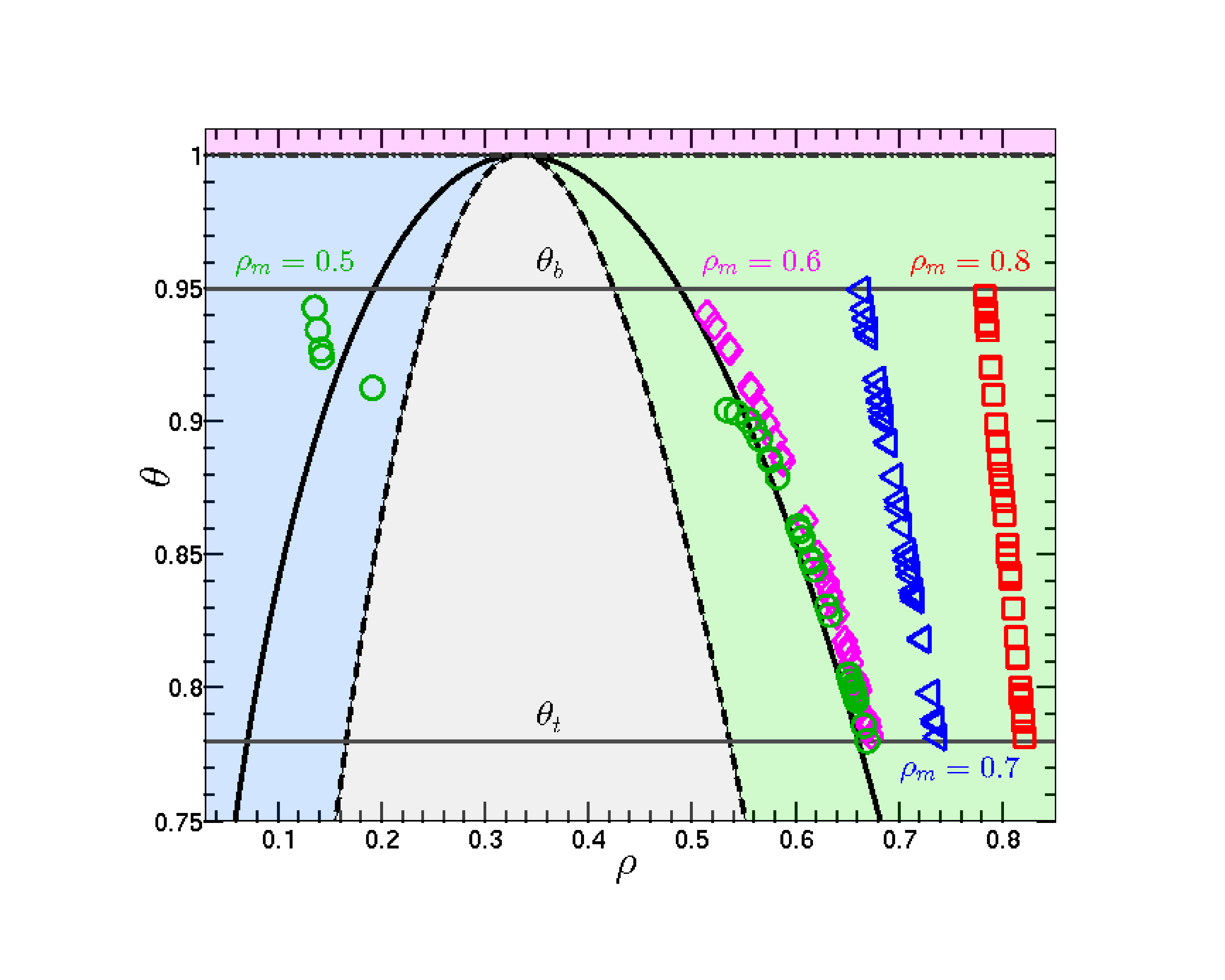} 
\end{tabular}
\caption{The black solid line is the coexistence line in the $\rho$-$\theta$ phase space; the black dashed line is the spinodal line; the solid grey horizontal lines correspond to the constant top and bottom temperatures. The blue shaded area represents the vapor state; the green shaded area represents the liquid state; the red shaded area represents the supercritical fluid state; the grey shaded area corresponds to the unstable elliptic region \cite{Johnston2014}. A scattered plot of the temperature and density values is superimposed. Their values are sampled at random spatial positions of the steady state solutions for $\rho_m = 0.8$ (red squares), $\rho_m = 0.7$ (blue triangles), $\rho_m = 0.6$ (magenta diamonds), and $\rho_m = 0.5$ (green circles).} 
\label{fig:sample_rho_theta}
\end{center}
\end{figure}

\section{Results}
\subsection{Solution at small Rayleigh number}
When the Rayleigh number is below the onset of instability, the system will evolve towards a steady state. The density and temperature profiles can be found by setting all time derivatives and $\mathbf u$ as zero in \eqref{eq:dimless_nsk_mass}-\eqref{eq:dimless_nsk_energy}:
\begin{eqnarray}
\label{eq:small_ra_eq_mom}
\nabla p - \nabla \cdot \boldsymbol \varsigma &= \rho \mathbf g, \\
\label{eq:small_ra_eq_ene}
\nabla \cdot \mathbf q &= 0.
\end{eqnarray}
In this example, the material moduli are chosen as $\bar{\mu} = 9.156 \times 10^{-3}$, $\kappa = 1.175 \times 10^{-2}$, $\lambda = 9.0\times 10^{-6}$, $\theta_m = 0.865$, and $\Delta \theta = 0.17$.
Notice that, instead of the thermal diffusivity $\alpha$, the thermal conductivity $\kappa$ is fixed as a constant in this example. Hence, the equation \eqref{eq:small_ra_eq_ene} is a linear Laplace's equation for the temperature field, and it can be solved analytically. The analytic steady state temperature profile is $\theta = \theta_t + \Delta \theta z / H$. Obtaining an analytic solution for the steady state density profile is non-trivial since it involves solving a third-order partial differential equation \eqref{eq:small_ra_eq_mom}. The initial density is homogeneous and $\rho_m = 0.8$, $0.7$, $0.6$, and $0.5$ respectively. The density profiles of the steady state solutions are illustrated in Fig. \ref{fig:smallRa_all}. All the steady state solutions show a stratification pattern with low-density fluid in the bottom and high-density fluid in the top. The difference between the four solutions can be better illustrated by sampling the density and temperature of the solutions at random spatial locations (Fig. \ref{fig:sample_rho_theta}). For $\rho_m = 0.8$, $0.7$, and $0.6$, all the sampled particles fall into the liquid state (the green shaded area in Fig. \ref{fig:sample_rho_theta}). Hence, the corresponding solutions shown in Fig. \ref{fig:smallRa_all} (a), (b), and (c) are all pure liquid with stratification. Unlike classical fluid stratification where the low-density fluid is above the high-density fluid, the Rayleigh number is small in these cases. The low Rayleigh number implies the relative strength of the gravity is small, and hence the fluid stratification is mainly driven by the temperature gradient. The linear temperature profile leads to a steady state pattern with light fluid in the bottom and dense fluid in the top. For $\rho_m = 0.5$, Fig. \ref{fig:sample_rho_theta} shows that a fraction of the fluid transits to the vapor state. Consequently, the result shown in Fig. \ref{fig:smallRa_all} (d) is different from the results in Fig. \ref{fig:smallRa_all} (a), (b) and (c). The steady state solution for $\rho_m = 0.5$ consists of separated liquid and vapor states and a thin transitional layer. The vapor state is located at the bottom of the domain and the liquid state is located on top of the vapor fluid. This configuration is similar to the steady state liquid-vapor two-phase solution obtained under zero gravity \cite{Onuki2007}.

\begin{figure}[htp!]
\begin{tabular}{cc}
\scalebox{0.09}{\includegraphics[angle=0, trim=60 170 30 50,
clip=true]{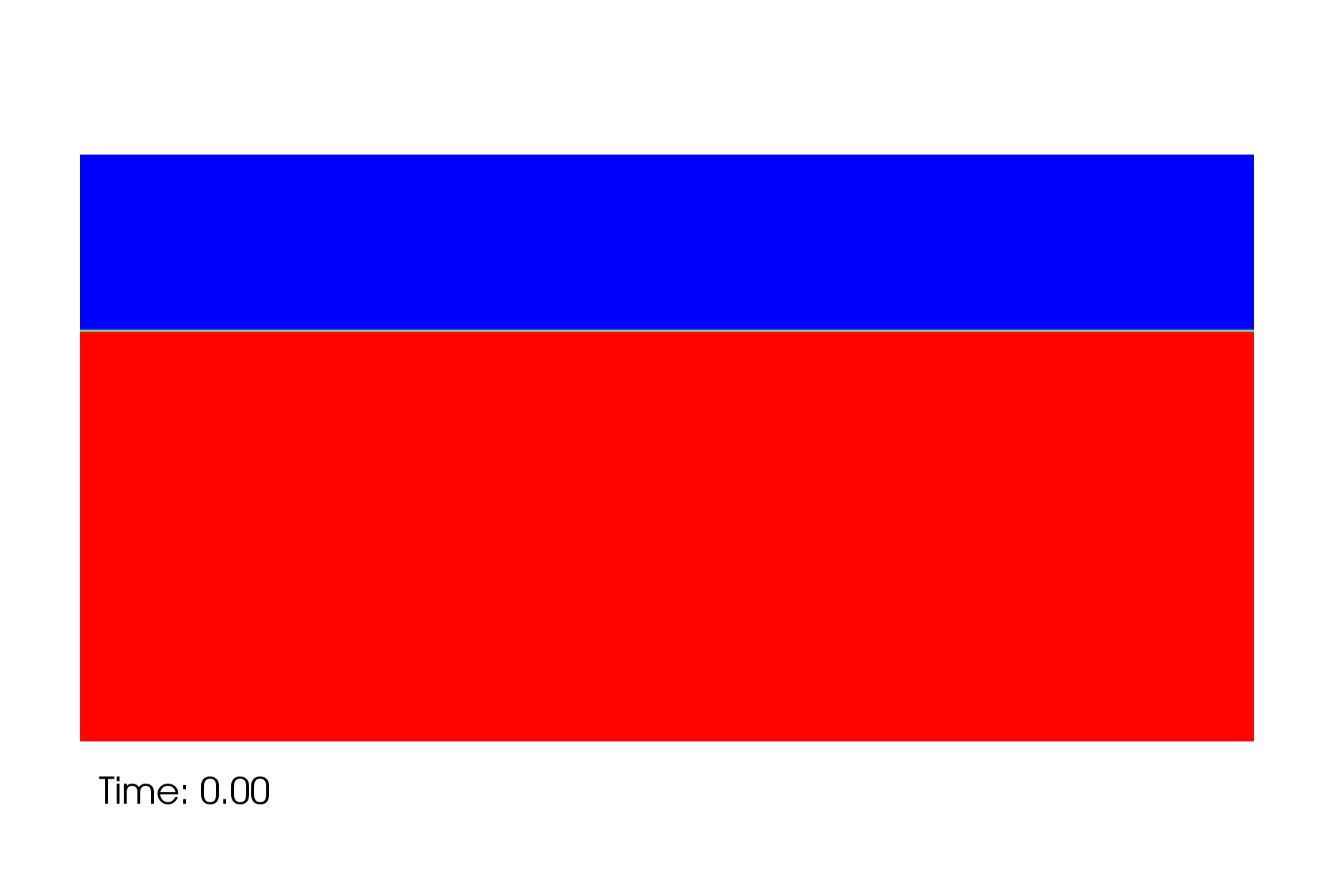} } & 
\scalebox{0.09}{\includegraphics[angle=0, trim=60 170 30 50,
clip=true]{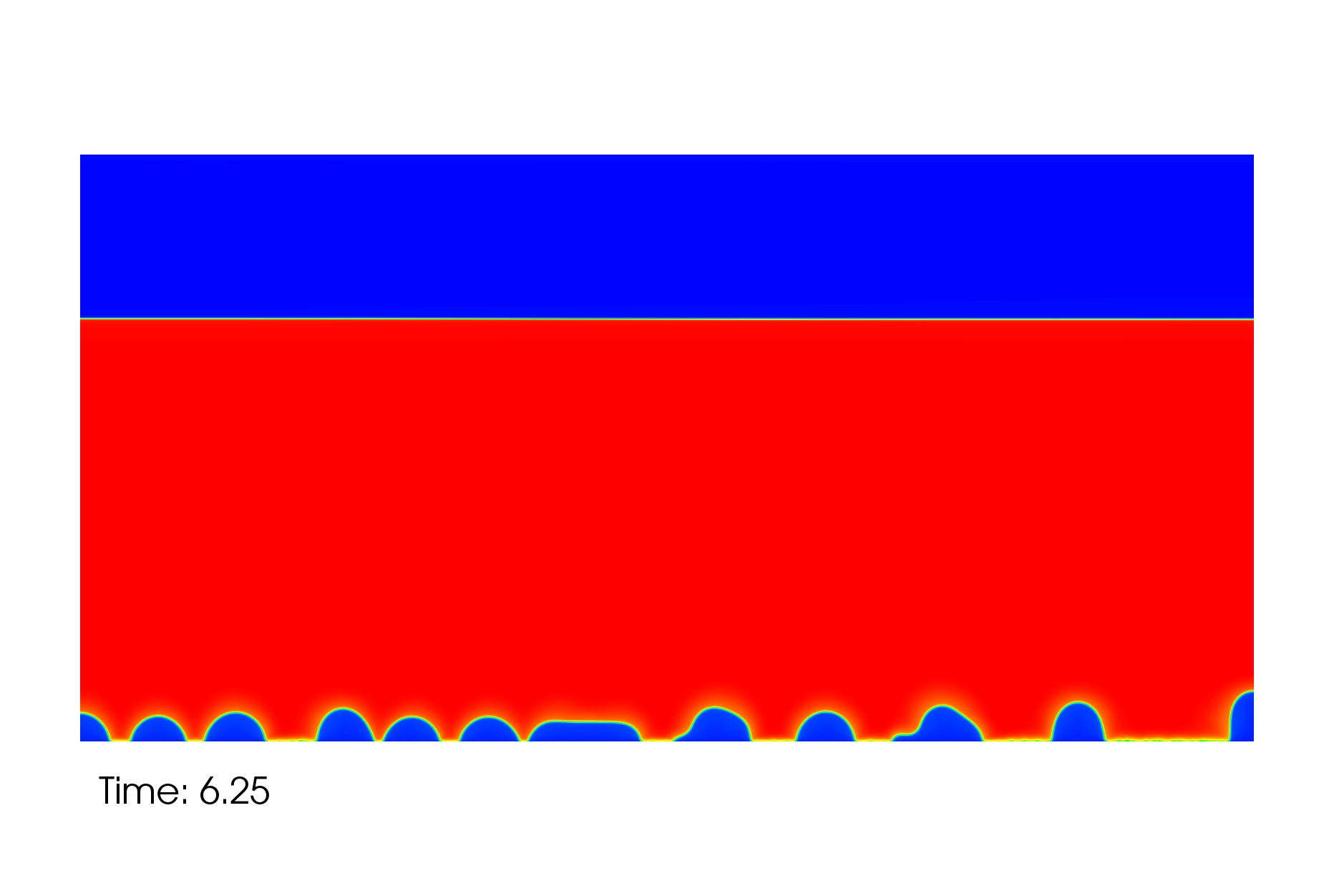} } \\
(a) & (b) \\
\scalebox{0.09}{\includegraphics[angle=0, trim=60 170 30 50,
clip=true]{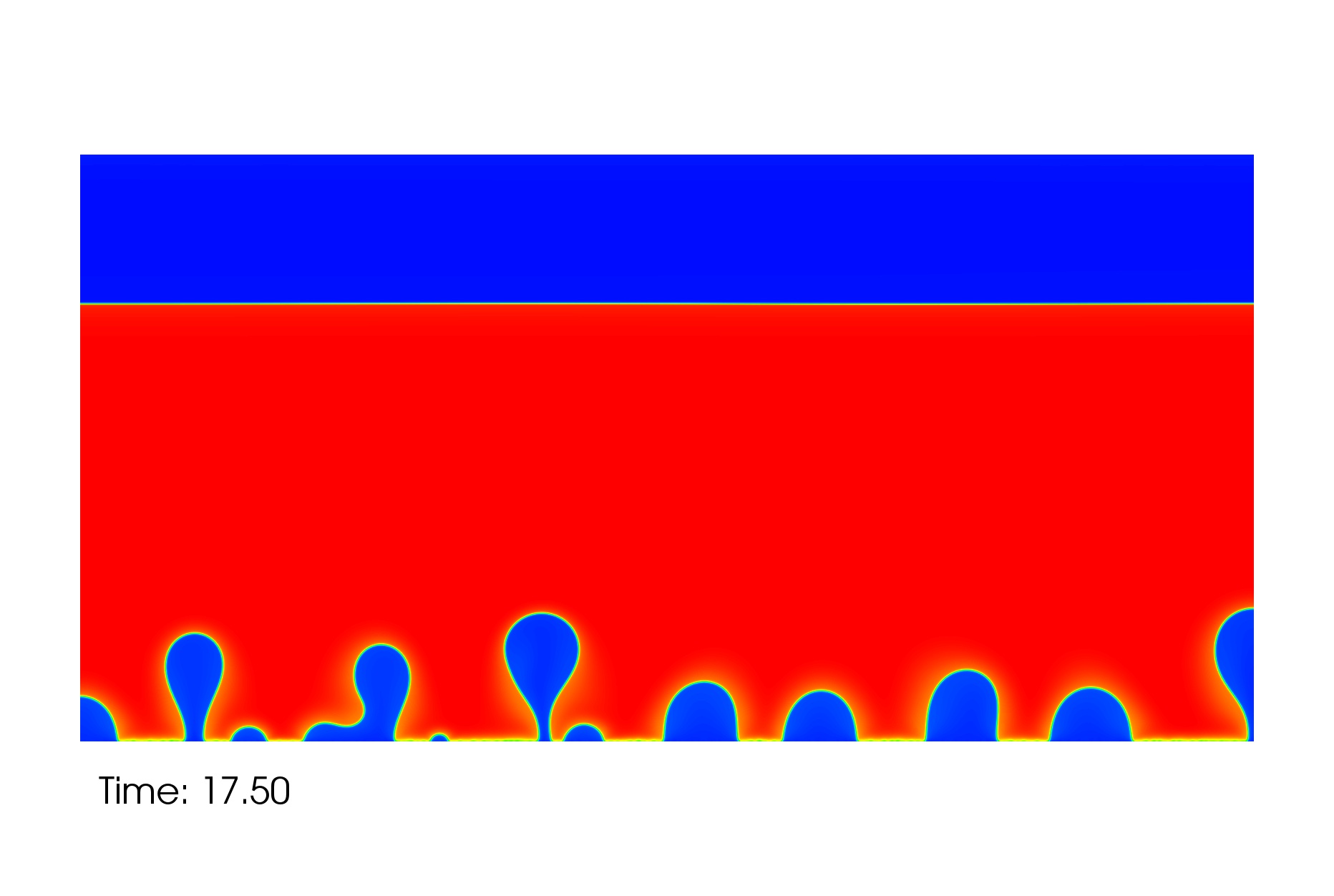} } & 
\scalebox{0.09}{\includegraphics[angle=0, trim=60 170 30 50,
clip=true]{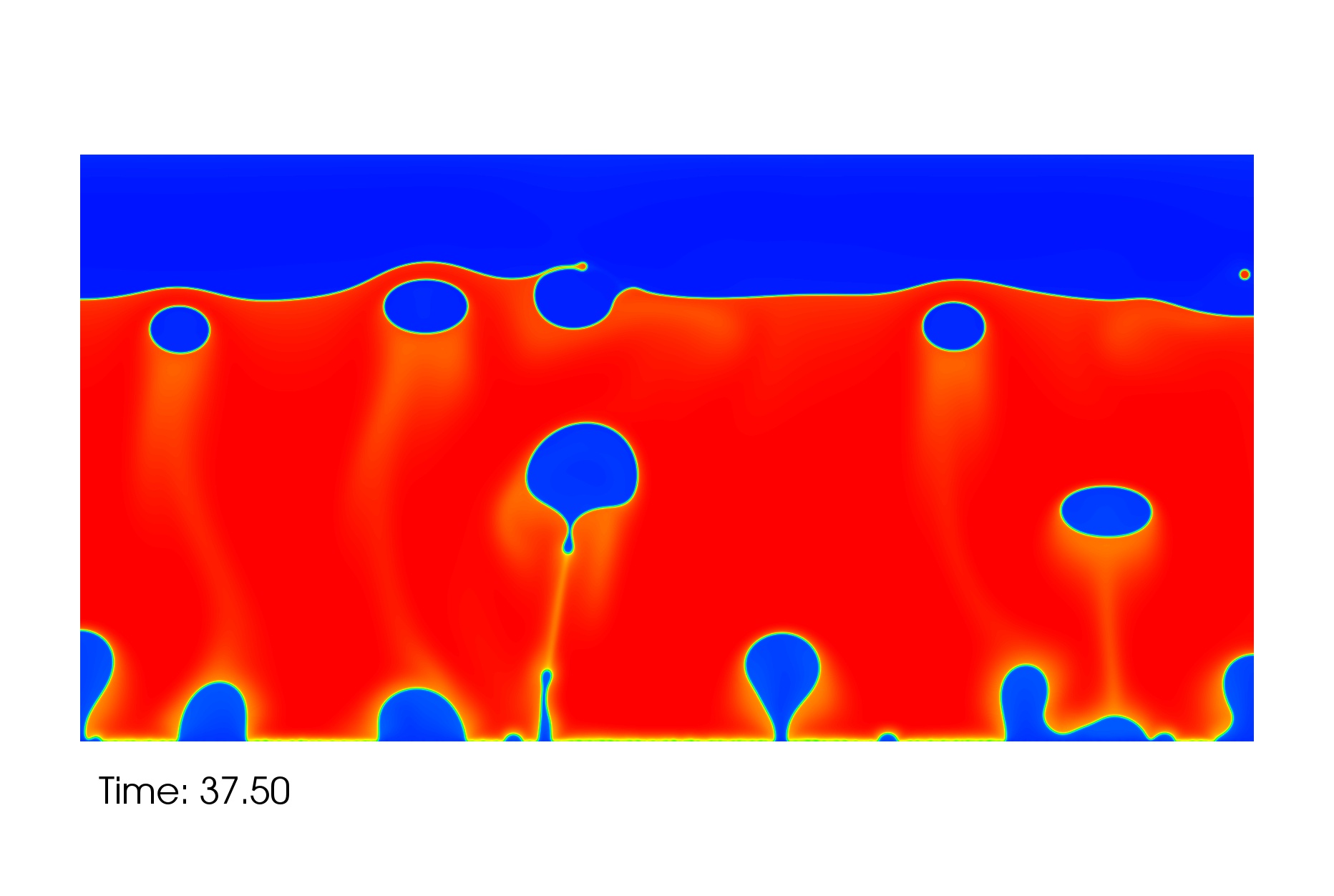} } \\
(c) & (d) \\
\scalebox{0.09}{\includegraphics[angle=0, trim=60 170 30 50,
clip=true]{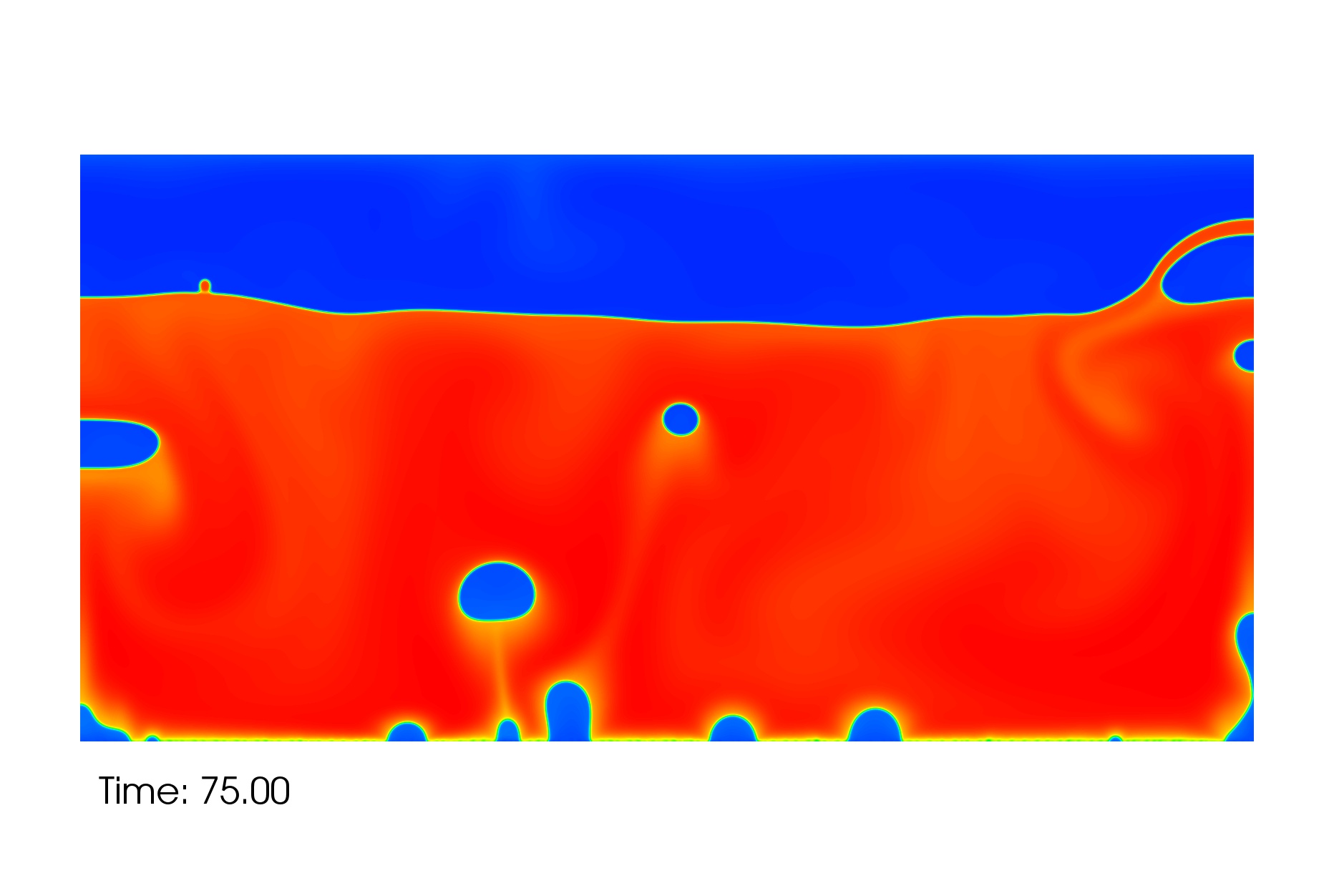} } & 
\scalebox{0.09}{\includegraphics[angle=0, trim=60 170 30 50,
clip=true]{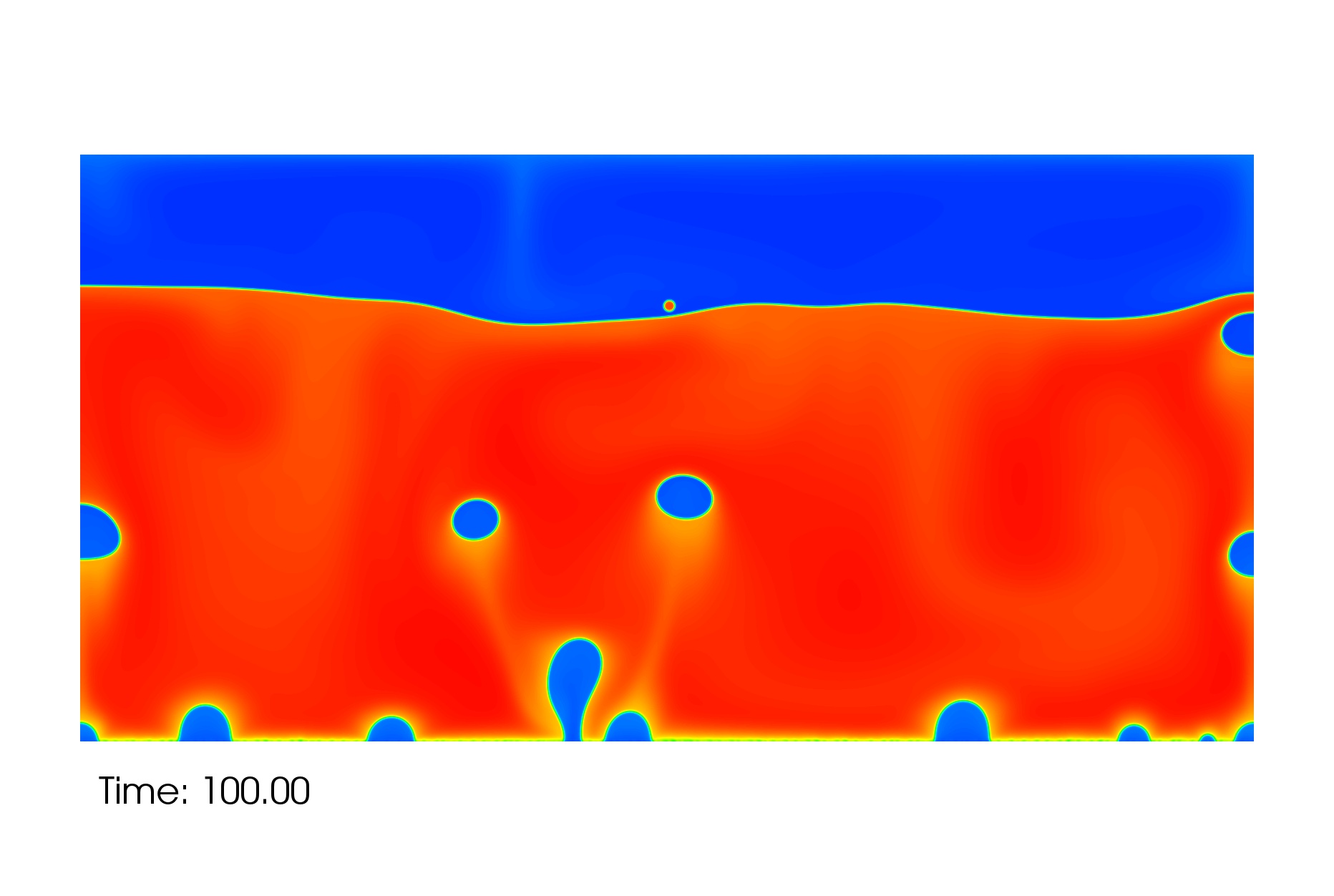} } \\
(e) & (f) 
\end{tabular}
\scalebox{0.19}{\includegraphics[angle=0, trim=90 900 90 200,
clip=true]{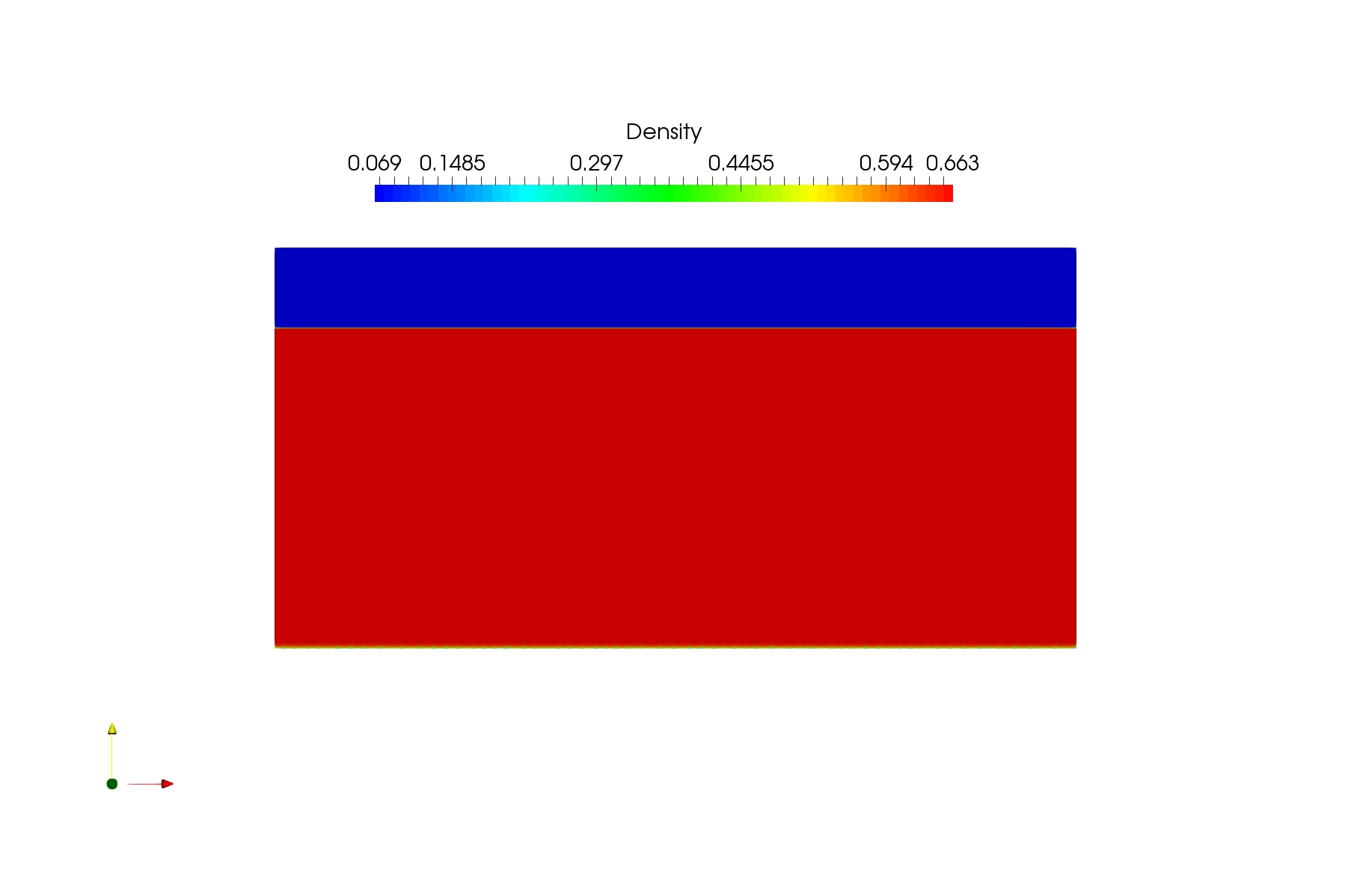} }
\centering
\caption{Two-dimensional nucleate boiling simulation: Density profiles at (a) $t=0.0$, (b) $t=6.25$, (c) $t=17.50$, (d) $t=37.50$, (e) $t=75.0$, and (f) $t=100.0$.}
\label{fig:boiling_2d_21}
\end{figure}

\begin{figure}[htp!]
\begin{tabular}{cc}
\scalebox{0.09}{\includegraphics[angle=0, trim=60 170 30 50,
clip=true]{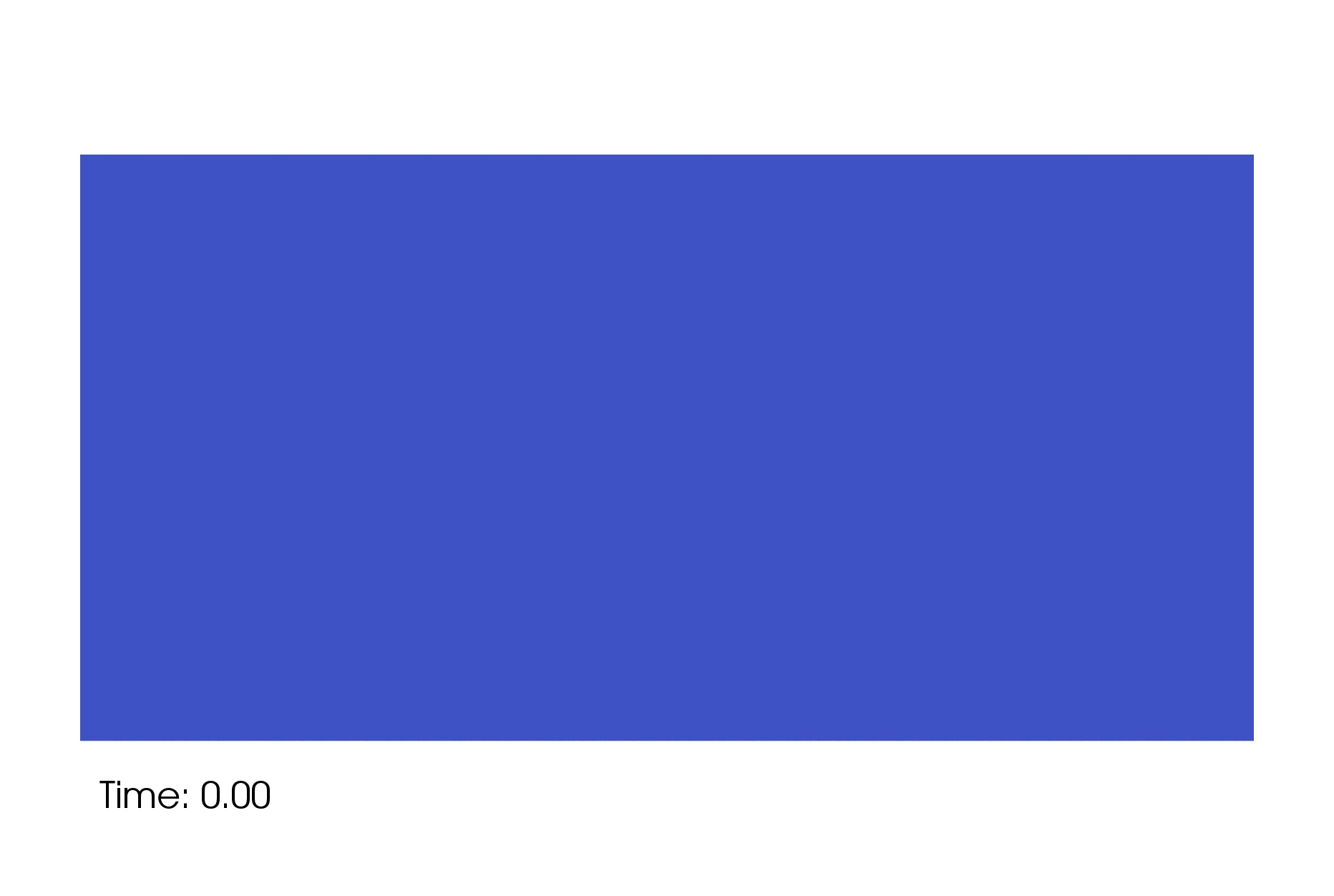} } & 
\scalebox{0.09}{\includegraphics[angle=0, trim=60 170 30 50,
clip=true]{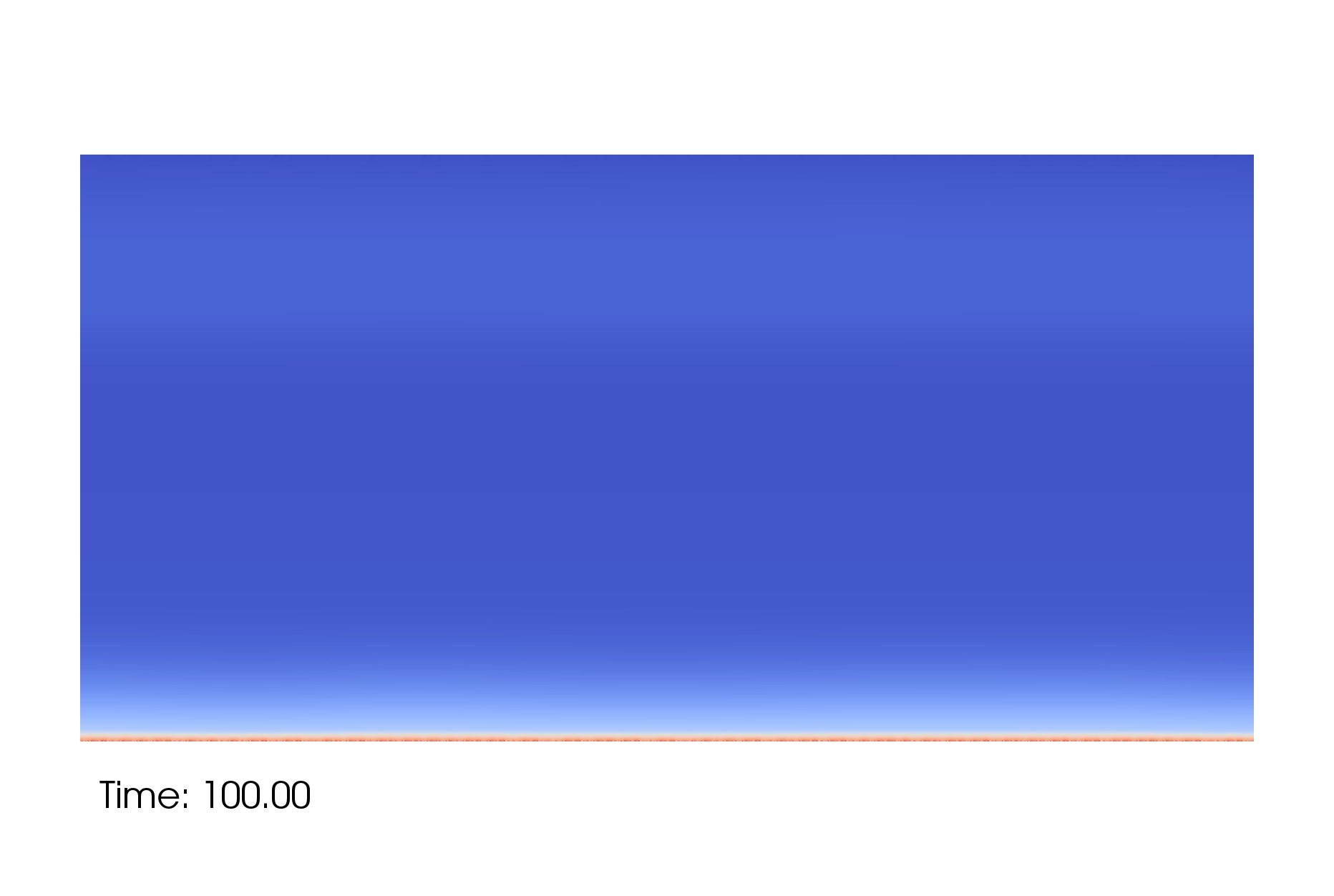} } \\
(a) & (b) \\
\scalebox{0.09}{\includegraphics[angle=0, trim=60 170 30 50,
clip=true]{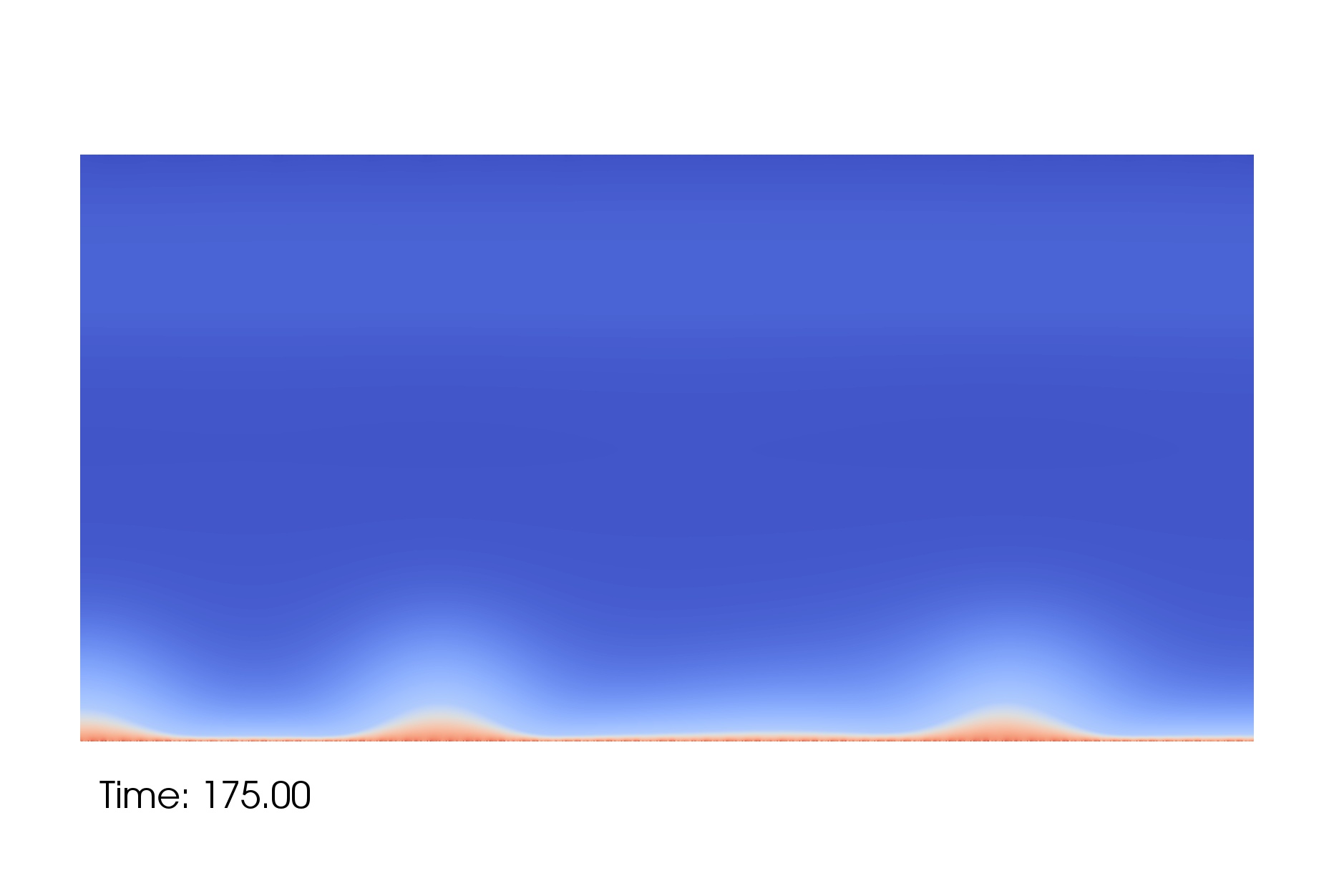} } & 
\scalebox{0.09}{\includegraphics[angle=0, trim=60 170 30 50,
clip=true]{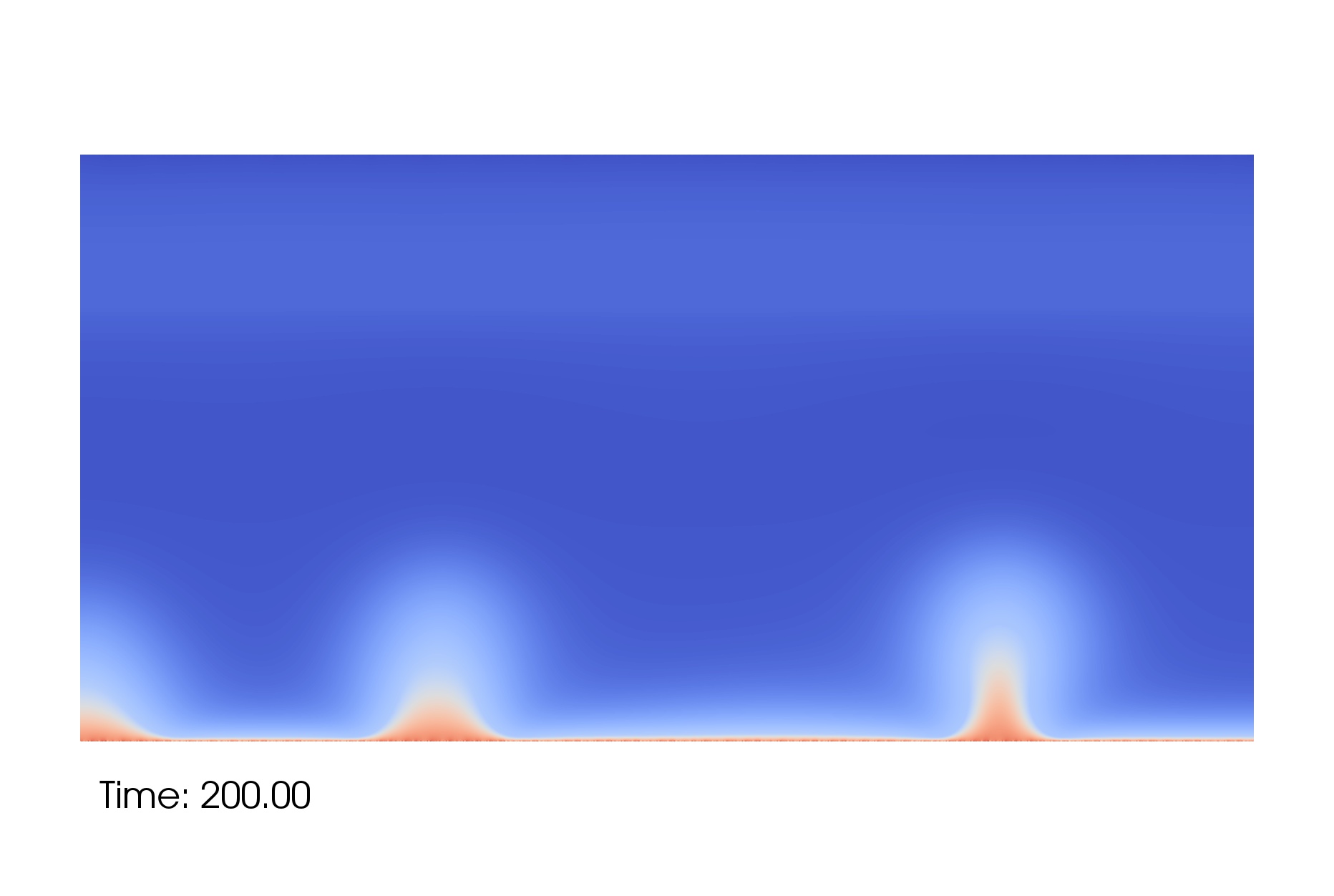} } \\
(c) & (d) \\
\scalebox{0.09}{\includegraphics[angle=0, trim=60 170 30 50,
clip=true]{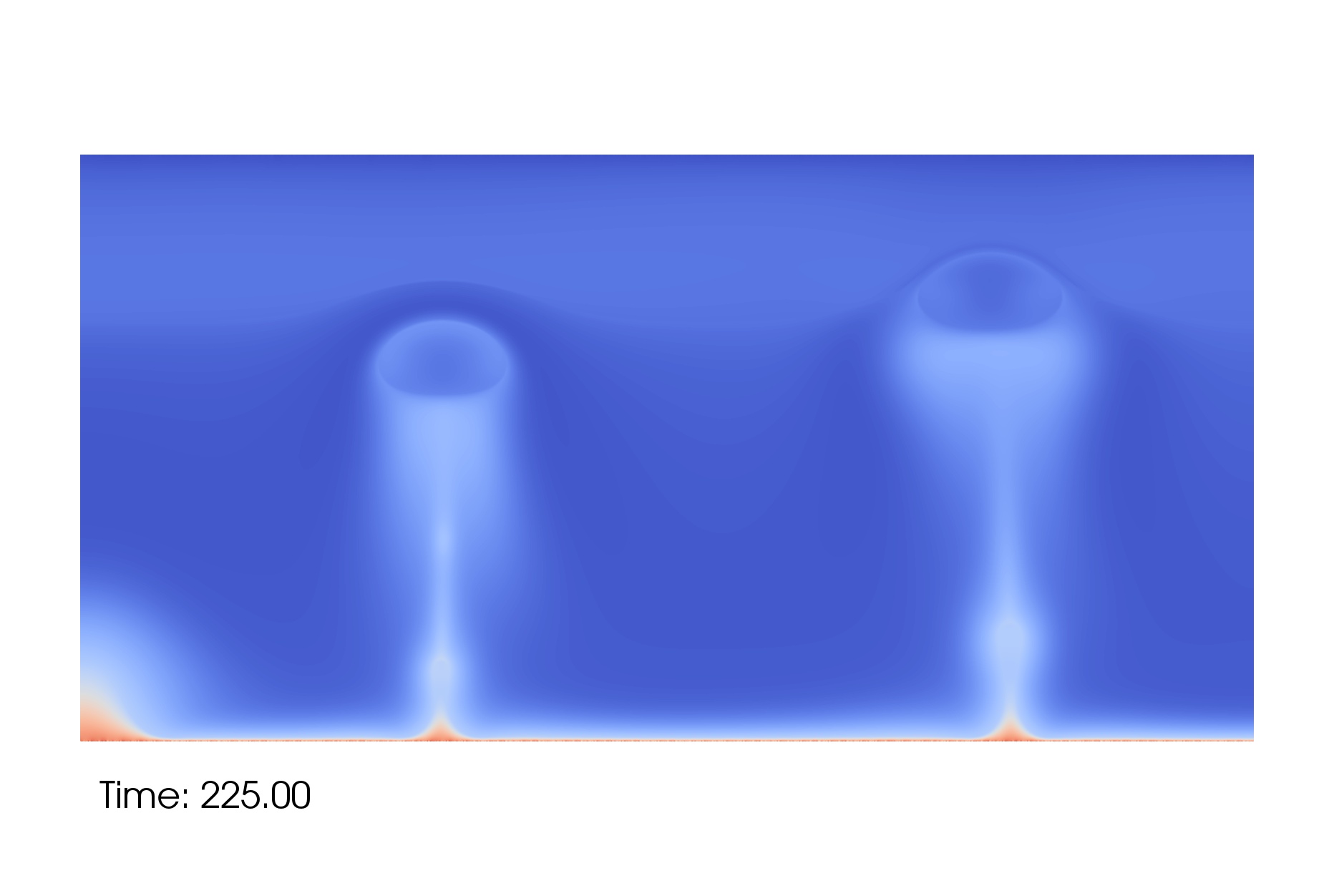} } & 
\scalebox{0.09}{\includegraphics[angle=0, trim=60 170 30 50,
clip=true]{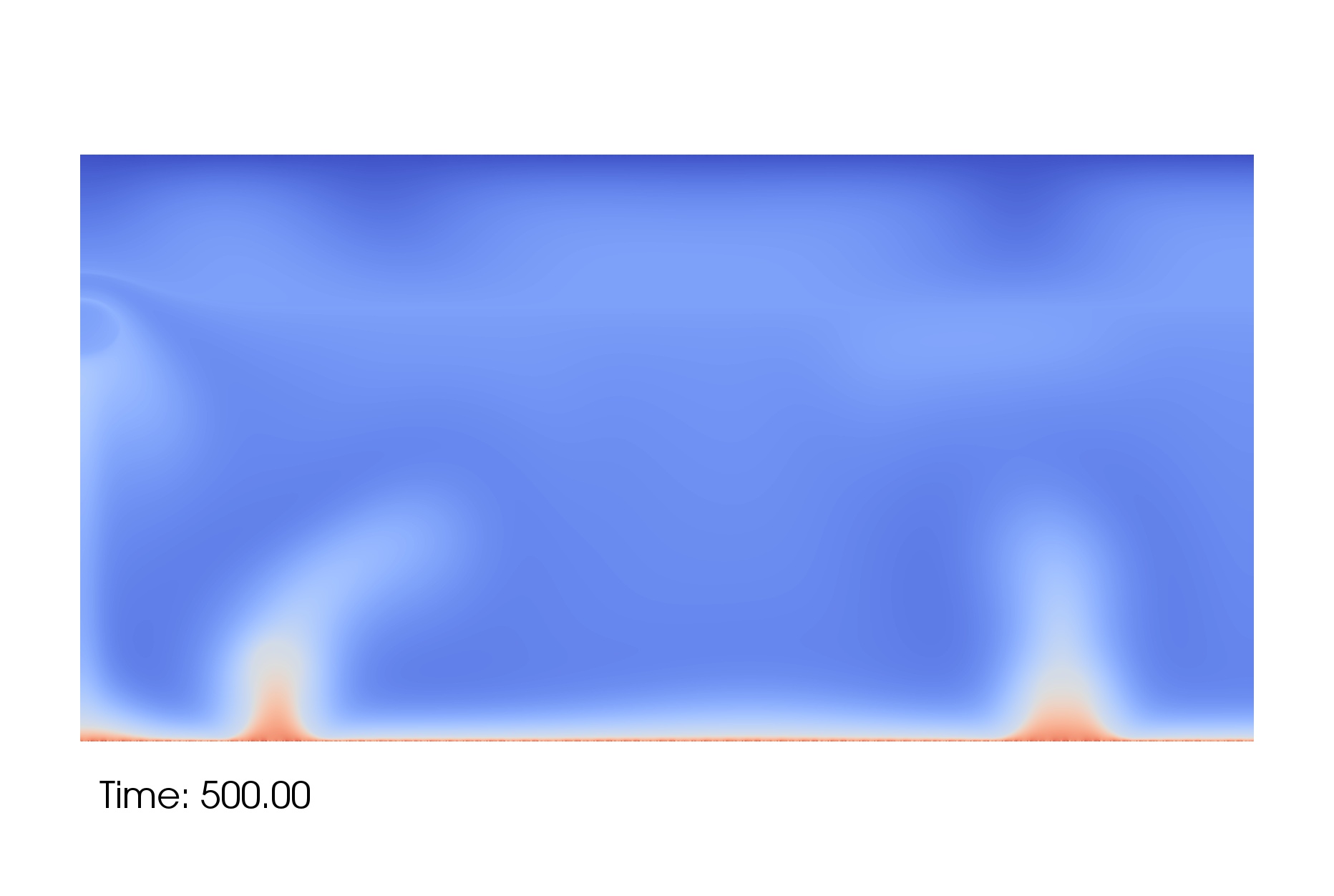} } \\
(e) & (f) 
\end{tabular}
\scalebox{0.19}{\includegraphics[angle=0, trim=90 900 90 180,
clip=true]{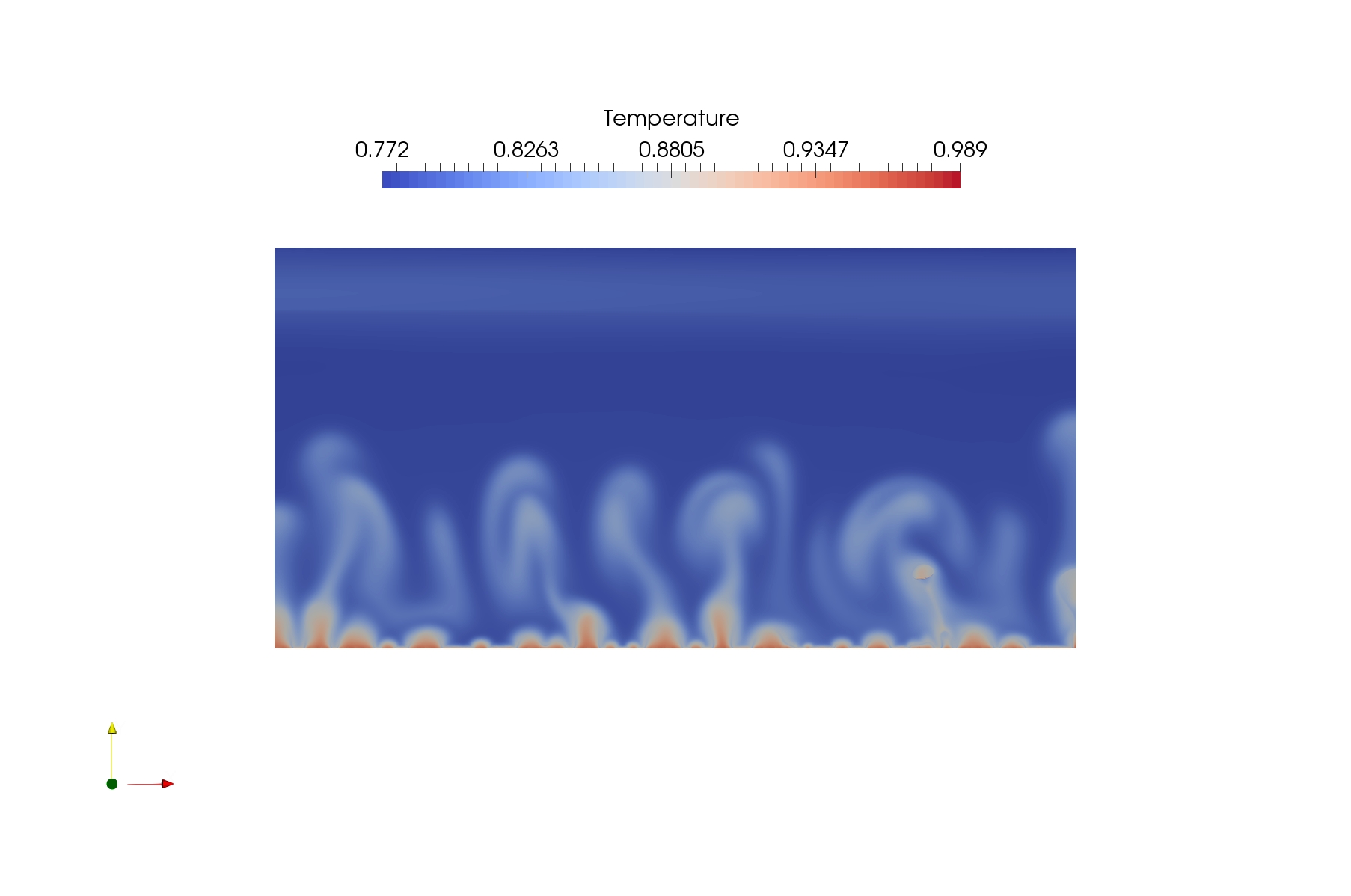} }
\centering
\caption{Two-dimensional film boiling simulation: Temperature profiles at (a) $t=0.0$, (b) $t=100.0$, (c) $t=175.0$, (d) $t=200.0$, (e) $t=225.0$, and (f) $t=500.0$.}
\label{fig:boiling_2d_23}
\end{figure}

\subsection{Nucleate and film boiling}
In this section, I numerically study the capability of the van der Waals fluid model in modeling different regimes of boiling. In the first simulation, parameters are chosen as $\bar{\nu} = 1.150 \times 10^{-4}$, $\alpha = 1.725 \times 10^{-5}$, $\lambda = 1.190 \times 10^{-7}$, $\theta_m = 0.8625$, $\Delta \theta = 0.175$, and $\rho_m = 0.2424$. The spatial domain is discretized by $2048 \times 1024$ quadratic B-splines and the time integration is performed up to $T=100.0$ with a fixed time step size $\Delta t = 5.0 \times 10^{-4}$. In Fig. \ref{fig:boiling_2d_21}, snapshots of the density profiles are illustrated at different time steps. The initial condition of this simulation represents the liquid fluid at the bottom and the vapor fluid at the top. A static free interface is located along $z=0.35$, and the initial temperature is 0.775. During the initial times, random small vapor bubbles are generated at the heated bottom surface and rise upward. At about $t=37.50$, the first a few bubbles reach the free surface, and in the meantime, there are more bubbles generated from the bottom surface. The coalescence of the vapor bubble with the free surface leads to surface waves. At $t=37.50$ and $t=75.0$, one can observe the surface waves. At time $t= 100.0$, there are tiny liquid droplets generated over the free surface as a result of the breakage of the liquid film.

In the next example, the kinematic viscosity is chosen as $4.600 \times 10^{-4}$, which is four times larger than that of the previous example. The increase in viscosity leads to slower dynamics of the fluid motion, and consequently, the numerical integration is performed up to $T=500.0$. All the other parameters are identical to those in the previous case. In Fig. \ref{fig:boiling_2d_23}, snapshots of the temperature at different time steps are depicted. It can be seen that during the initial times, there is a thin film generated at the bottom heated surface. As time evolves, the film becomes unstable and three bubbles are formed. The bubbles gradually get detached from the film and rise upward carrying heat away from the thin film. At the final state, there is a mixing pattern of the temperature field driven by the free convection.

\begin{figure}[htp!]
\begin{center}
\begin{tabular}{cc}
\scalebox{0.3}{\includegraphics[angle=0, trim=30 530 30 50,
clip=true]{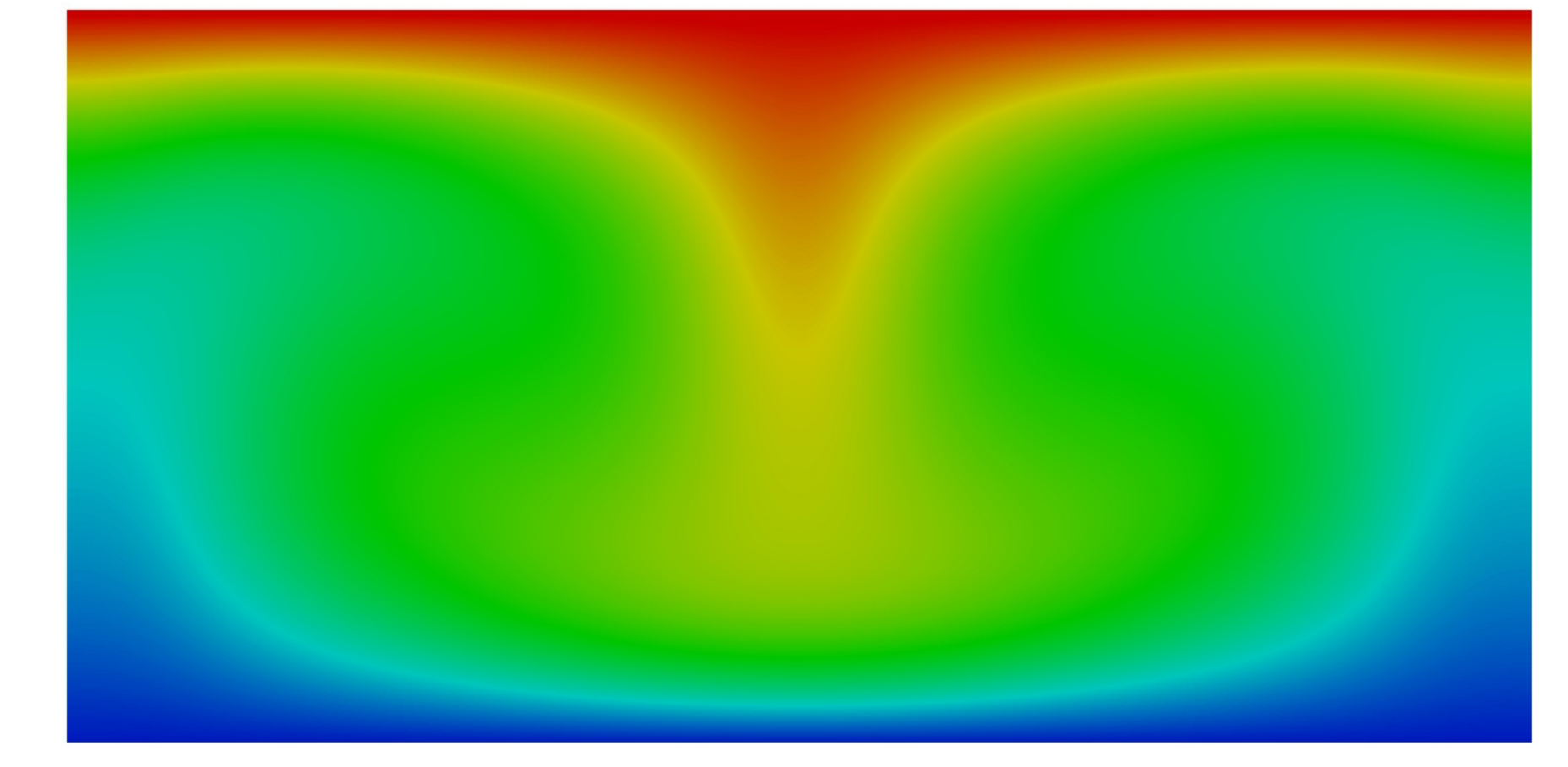} } &
\scalebox{0.3}{\includegraphics[angle=0, trim=30 530 30 50,
clip=true]{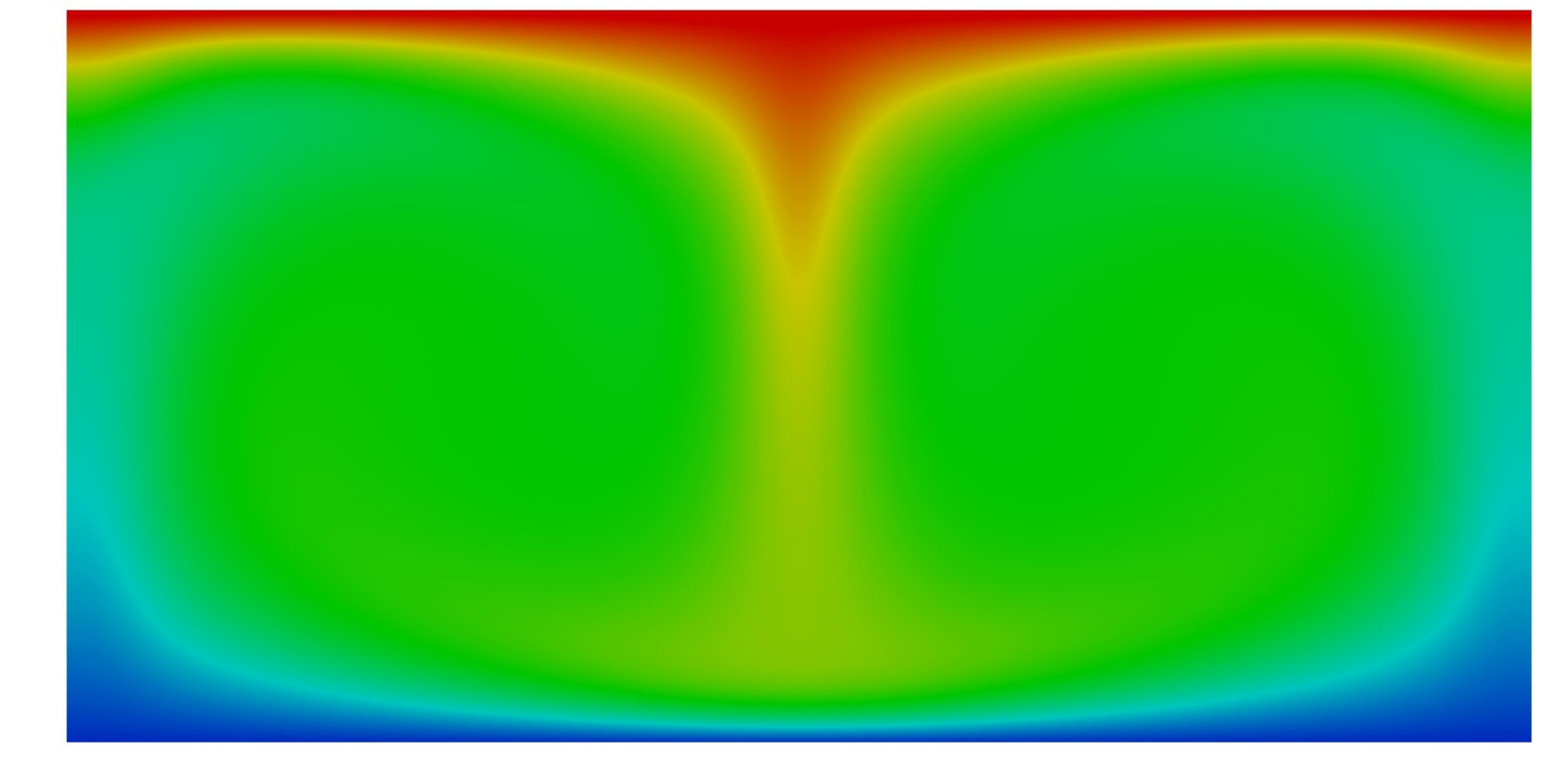} } \\
(a) & (b) \\
\scalebox{0.3}{\includegraphics[angle=0, trim=30 530 30 50,
clip=true]{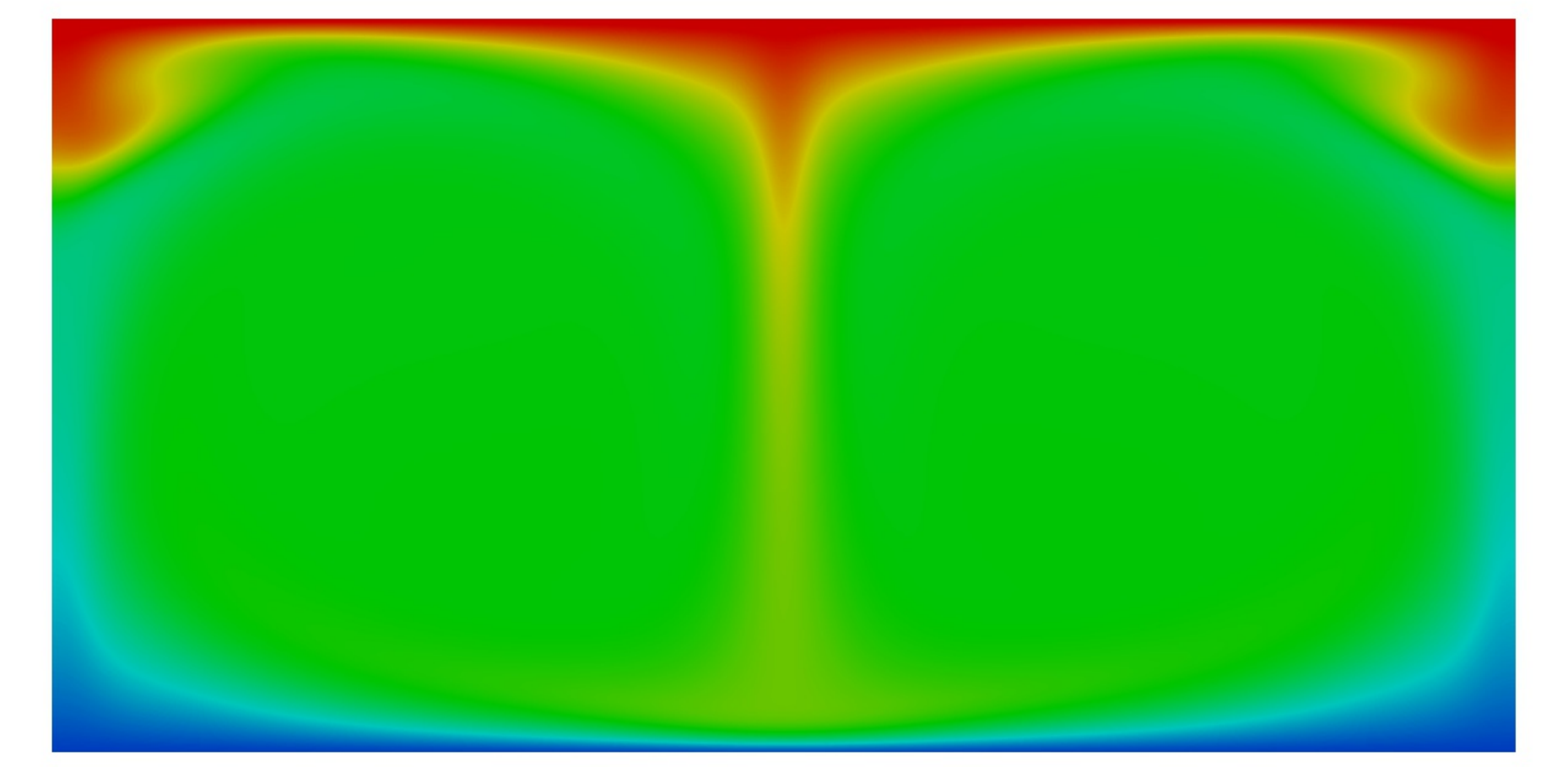} } &
\scalebox{0.3}{\includegraphics[angle=0, trim=30 530 30 50,
clip=true]{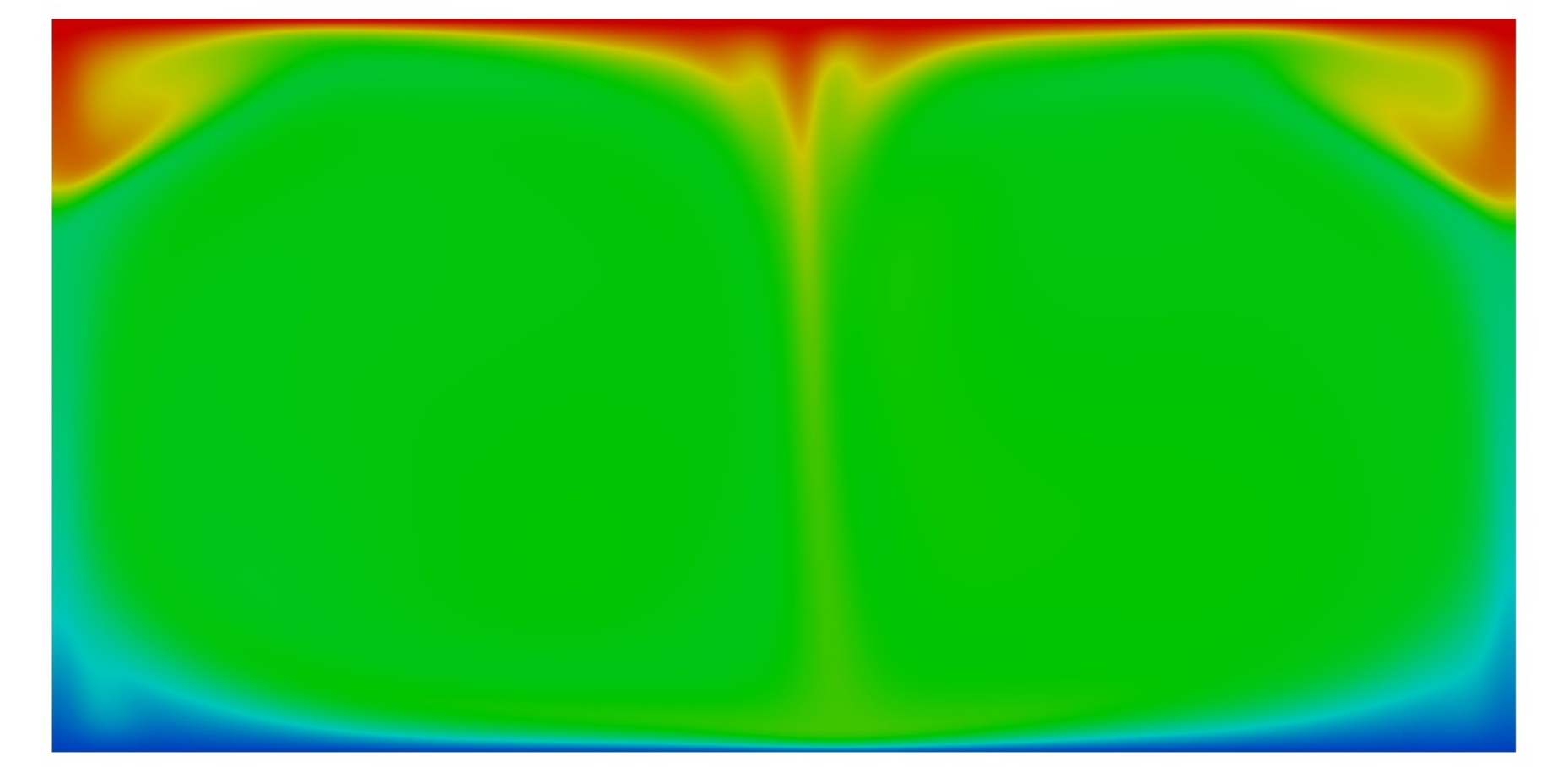} } \\
(c) & (d)
\end{tabular}
\scalebox{0.45}{\includegraphics[angle=0, trim=90 700 90 80,
clip=true]{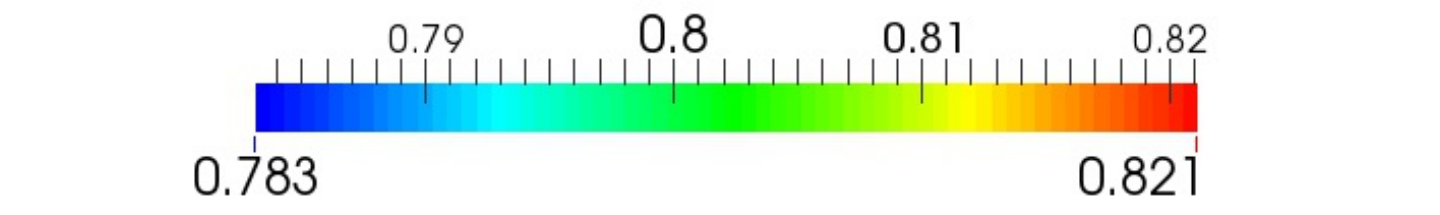} }
\caption{The density profiles of the solutions at time step $t=5000.0$: (a) $\operatorname{Ra} = 1.0\times 10^5$, (b) $\operatorname{Ra} = 1.0\times 10^6$, (c) $\operatorname{Ra} = 1.0\times 10^7$, and (d) $\operatorname{Ra} = 1.0\times 10^8$. }
\label{fig:Rho_0d8_rho}
\end{center}
\end{figure}

\begin{figure}[htp!]
\begin{center}
\begin{tabular}{cc}
\scalebox{0.3}{\includegraphics[angle=0, trim=30 530 30 50,
clip=true]{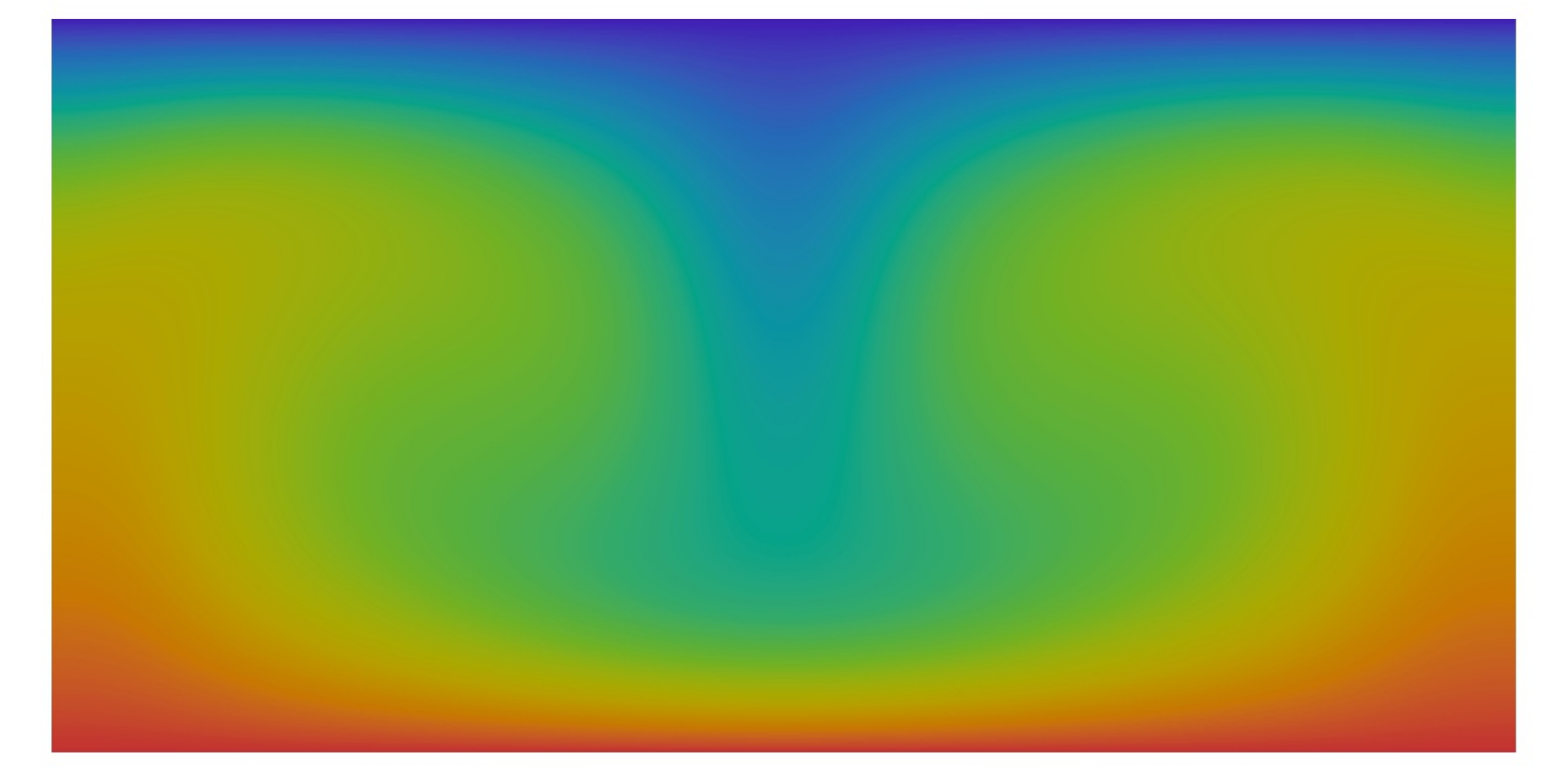} } &
\scalebox{0.3}{\includegraphics[angle=0, trim=30 530 30 50,
clip=true]{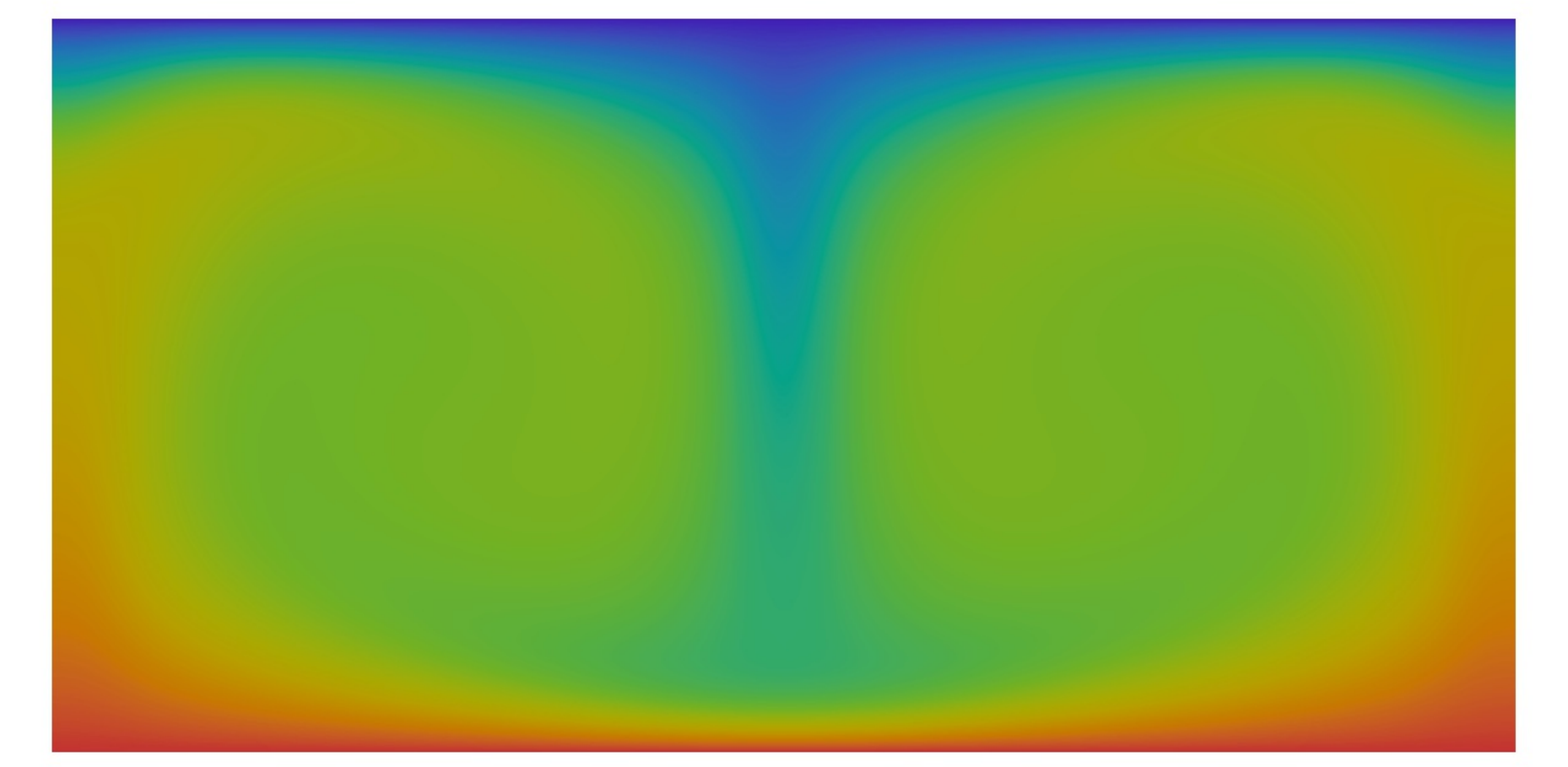} } \\
(a) & (b) \\
\scalebox{0.3}{\includegraphics[angle=0, trim=30 530 30 50,
clip=true]{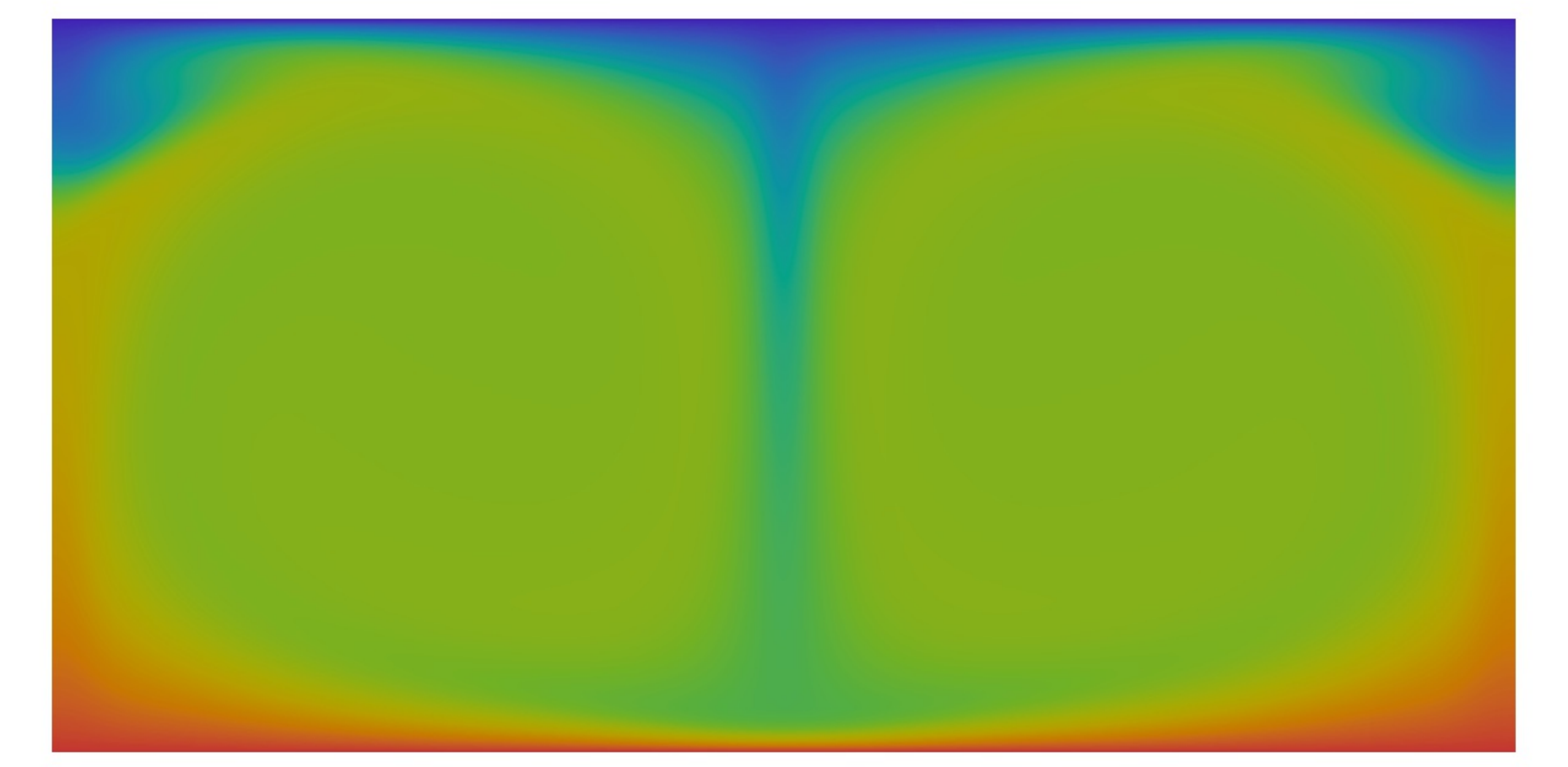} } &
\scalebox{0.3}{\includegraphics[angle=0, trim=30 530 30 50,
clip=true]{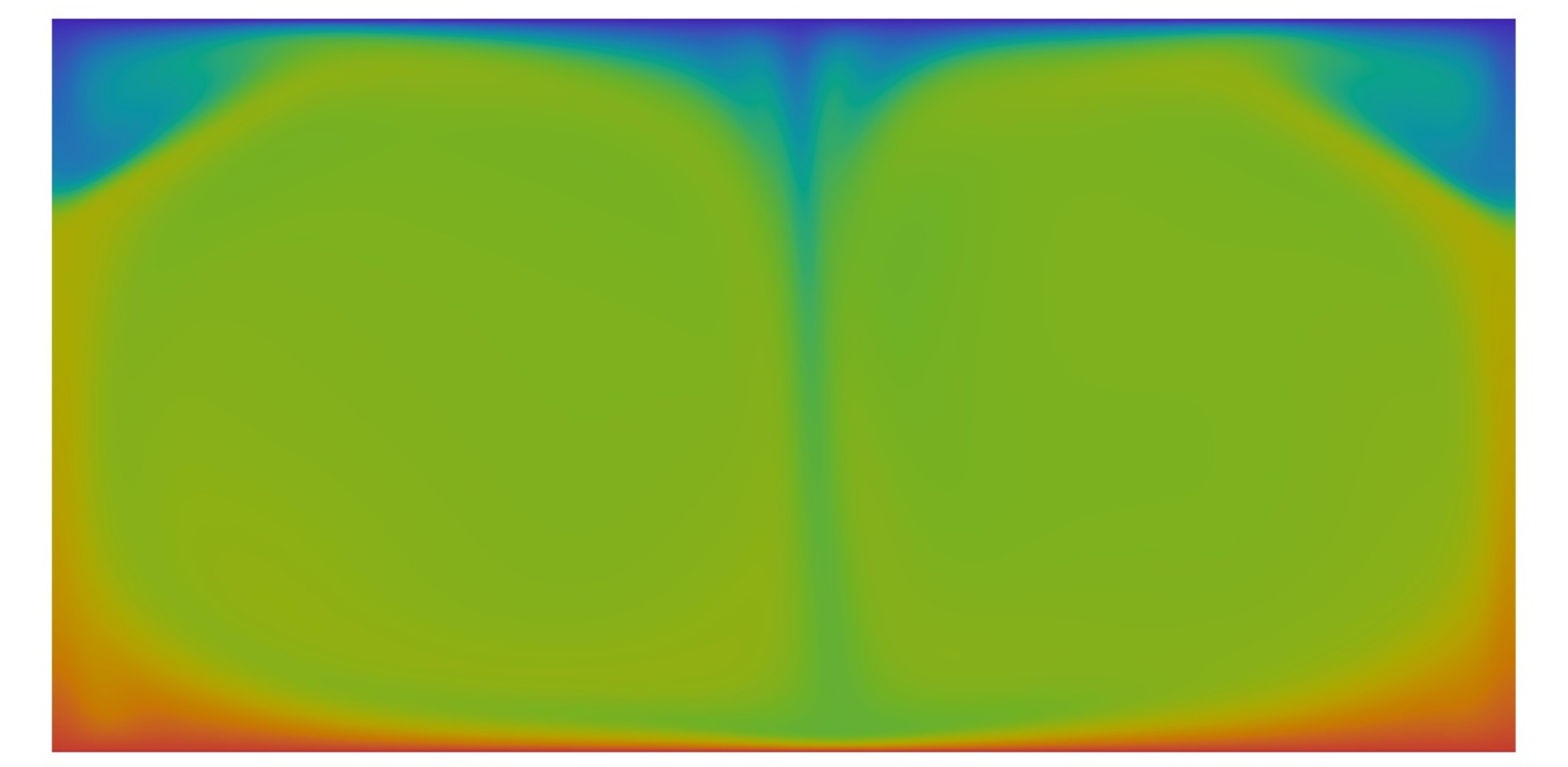} } \\
(c) & (d)
\end{tabular}
\scalebox{0.45}{\includegraphics[angle=0, trim=90 700 90 80,
clip=true]{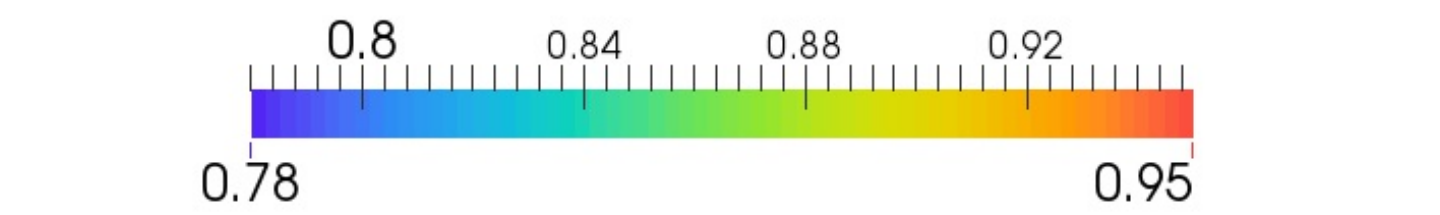} }
\caption{The temperature profiles of the solutions at time step $t=5000.0$: (a) $\operatorname{Ra} = 1.0\times 10^5$, (b) $\operatorname{Ra} = 1.0\times 10^6$, (c) $\operatorname{Ra} = 1.0\times 10^7$, and (d) $\operatorname{Ra} = 1.0\times 10^8$. }
\label{fig:Rho_0d8_temp}
\end{center}
\end{figure}

\begin{table}[htbp]
\begin{center}
\tabcolsep=0.19cm
\renewcommand{\arraystretch}{1.5}
\begin{tabular}{c c c c c c}
\hline
\hline
$N_x \times N_z$ & $\bar{\mu}$ & $\kappa$  & $\operatorname{Ra}(\rho_m, \theta_m)$ & $\operatorname{Pr}(\rho_m, \theta_m)$ & $\operatorname{Nu}_{V,t}$ \\
\hline
256$\times$128 & 9.156$\times 10^{-3}$ & 1.175$\times 10^{-2}$ & $1.0 \times 10^{1}$ & 1.0 & 1.000 \\
512$\times$256 & 9.156$\times 10^{-3}$ & 1.175$\times 10^{-2}$ & $1.0 \times 10^{1}$ & 1.0 & 1.000 \\
256$\times$128 & 9.156$\times 10^{-5}$ & 1.175$\times 10^{-4}$ & $1.0 \times 10^{5}$ & 1.0 & 2.921 \\
512$\times$256 & 9.156$\times 10^{-5}$ & 1.175$\times 10^{-4}$ & $1.0 \times 10^{5}$ & 1.0 & 2.924 \\
256$\times$128 & 4.095$\times 10^{-5}$ & 5.257$\times 10^{-5}$ & $5.0 \times 10^{5}$ & 1.0 & 4.608 \\
512$\times$256 & 4.095$\times 10^{-5}$ & 5.257$\times 10^{-5}$ & $5.0 \times 10^{5}$ & 1.0 & 4.619 \\
256$\times$128 & 2.895$\times 10^{-5}$ & 3.717$\times 10^{-5}$ & $1.0 \times 10^{6}$ & 1.0 & 5.532 \\
512$\times$256 & 2.895$\times 10^{-5}$ & 3.717$\times 10^{-5}$ & $1.0 \times 10^{6}$ & 1.0 & 5.546 \\
256$\times$128 & 1.295$\times 10^{-5}$ & 1.662$\times 10^{-5}$ & $5.0 \times 10^{6}$ & 1.0 & 8.032 \\
512$\times$256 & 1.295$\times 10^{-5}$ & 1.662$\times 10^{-5}$ & $5.0 \times 10^{6}$ & 1.0 & 8.032 \\
256$\times$128 & 9.156$\times 10^{-6}$ & 1.175$\times 10^{-5}$ & $1.0 \times 10^{7}$ & 1.0 & 9.236 \\
512$\times$256 & 9.156$\times 10^{-6}$ & 1.175$\times 10^{-5}$ & $1.0 \times 10^{7}$ & 1.0 & 9.311 \\
256$\times$128 & 4.095$\times 10^{-6}$ & 5.257$\times 10^{-6}$ & $5.0 \times 10^{7}$ & 1.0 & 12.187 \\
512$\times$256 & 4.095$\times 10^{-6}$ & 5.257$\times 10^{-6}$ & $5.0 \times 10^{7}$ & 1.0 & 12.582 \\
256$\times$128 & 2.895$\times 10^{-6}$ & 3.717$\times 10^{-6}$ & $1.0 \times 10^{8}$ & 1.0 & 14.264 \\
512$\times$256 & 2.895$\times 10^{-6}$ & 3.717$\times 10^{-6}$ & $1.0 \times 10^{8}$ & 1.0 & 14.653 \\
512$\times$256 & 1.295$\times 10^{-6}$ & 1.662$\times 10^{-6}$ & $5.0 \times 10^{8}$ & 1.0 & 19.952 \\
512$\times$256 & 9.156$\times 10^{-7}$ & 1.175$\times 10^{-6}$ & $1.0 \times 10^{9}$ & 1.0 & 22.305 \\
\hline 
\hline
\end{tabular}
\end{center}
\caption{Summary of the simulation results with $\lambda = 9.0\times 10^{-6}$, $\rho_m = 0.8$, $\theta_b = 0.95$, and $\theta_t = 0.78$. For the averaged density $\rho_m$ and the arithmetic mean temperature $\theta_m$, $c_{\beta}(\rho_m,\theta_m) = 0.3082$, $\chi(\rho_m, \theta_m)=1.3328$. Consequently, $\operatorname{Ra}(\rho_m, \theta_m) = 1.0762\times 10^{-3} / (\bar{\mu} \kappa) $ and $\operatorname{Pr}(\rho_m,\theta_m) = 1.2838\bar{\mu}/\kappa$.}
\label{table:tnsk_nusselt_rho0d8_we9e6}
\end{table}

\subsection{The Nusselt number scaling}
In this section, the flow structure of the free convection in the van der Waals fluid is explored by relating the Nusselt number with the Rayleigh number $\operatorname{Ra}$ at a fixed Prandtl number $\operatorname{Pr}$. In this suite of simulations, the volume averaged density $\rho_m$ is chosen as $0.8$. The viscosity coefficient $\bar{\mu}$ and the conductivity $\kappa$ are progressively reduced so that the Rayleigh number ranges from $10$ to $1.0\times 10^9$ while the Prandtl number is maintained at $1.0$. For $\operatorname{Ra} \leq 10^8$, two different meshes are used for the same set of parameters to guarantee converged results for the Nusselt number. As is revealed in Fig. \ref{fig:sample_rho_theta}, the fluid in this simulation remain in the liquid phase, and the fluid density is stratified due to the temperature variation and the gravity force. In Fig. \ref{fig:Rho_0d8_rho} and \ref{fig:Rho_0d8_temp}, the density and the temperature are depicted at time $t=5000.0$ for different Rayleigh numbers. At smaller Rayleigh numbers, there are two convective rolls formed in the domain. For Rayleigh number $1.0\times 10^8$, the rolls become unstable and the symmetry is broken, as is shown in Fig. \ref{fig:Rho_0d8_temp} (d). The simulation results are reported in Table \ref{table:tnsk_nusselt_rho0d8_we9e6}. In this set of simulations, the value of $\gamma_{\textup{Nu}}$ is about $0.22$ by least square fitting. In \cite{Tilgner2011}, the value of $\gamma_{\textup{Nu}}$ for ideal gas is reported to be 0.265; in \cite{Lakkaraju2013}, the value of $\gamma_{\textup{Nu}}$ for bubbling flows is given to be between $1/5$ and $1/3$. This suggests that the numerically measured $\gamma_{\textup{Nu}}$ in the van der Waals fluid conforms to reported values.


\section{Conclusion}
In this work, the Rayleigh-B\'enard free convection in the van der Waals fluid is numerically investigated. Dimensional analysis is performed for the governing equations, and the control parameters for the convection problem are identified. The provably entropy stable algorithm and isogeometric analysis provide a reliable high-resolution numerical method for studying the free convection problem. The numerical results demonstrate the capability of the numerical model in describing boiling in different regimes. A suite of two-dimensional numerical simulations is also performed as an investigation of the heat transport property of the van der Waals fluid. The preliminary results indicate that this work provides a suitable framework to study the heat transport property for nucleate and film boiling. As the future work, the Nusselt number scaling law will be further investigated under different choices of the control parameters.

\section*{Acknowledgement}
The author acknowledges the Texas Advanced Computing Center (TACC) at The University of Texas at Austin for providing HPC, visualization, and data storage resources that have contributed to the results reported within this work.

\bibliographystyle{plain}
\bibliography{tnskbib}
\end{document}